\documentclass[twocolumn,showpacs,reprint,aps]{revtex4-1}

\usepackage{graphicx}
\usepackage{dcolumn}
\usepackage{bm}
\usepackage{epsfig}
\usepackage{longtable}

\newlength\figwidth\figwidth=0.5\textwidth
\renewcommand{\thefootnote}{\fnsymbol{footnote}}

\begin{document}

\title{$0^+$ states and collective bands in $^{228}$Th by  the (p,t) reaction}

\author{A.~I.~Levon$^{1}$,
 G.~Graw$^2$, R.~Hertenberger$^2$, S.~Pascu$^{3}$,
 P.~G.~Thirolf$^2$,  H.-F.~Wirth$^2$, and P.~Alexa$^4$}

\affiliation{$^1$ Institute for Nuclear Research, Academy of Science, Kiev, Ukraine}
\email[Electronic address: ] {levon@kinr.kiev.ua}

\affiliation{$^2$ Fakult\"at f\"ur Physik, Ludwig-Maximilians-Universit\"at M\"unchen, Garching,
Germany}

\affiliation{$^3$ Institut f\"ur Kernphysik, Universit\"at zu K\"oln,  K\"oln, Germany}
\altaffiliation [Permanent address: ] {H.~Hulubei National Institute of Physics and Nuclear
Engineering, Bucharest, Romania}

\affiliation{$^4$ Institute of Physics and Institute of Clean Technologies,  Technical
University of Ostrava, Czech Republic}

\begin{abstract}

 The excitation spectra in the deformed nucleus $^{228}$Th have been studied by means of
the (p,t) reaction, using the Q3D spectrograph facility at the Munich Tandem accelerator. The
angular distributions of tritons
 were measured for about 110 excitations seen in the triton spectra up
 to 2.5 MeV.  Firm $0^+$ assignments are made for 17 excited states by
 comparison of experimental angular distributions with the calculated
 ones using the CHUCK3 code.
 Assignments up to spin $6^+$  are made for other states. Sequences of  states
 are selected which can be treated as rotational bands and as multiplets
of excitations. Moments of inertia have been derived from these sequences, whose values may be
considered as evidence of the  two-phonon nature of most $0^+$ excitations. Experimental
 data are compared with  interacting boson model (IBM)  and
 quasiparticle-phonon model (QPM) calculations and with experimental data for $^{229}$Pa.

\end{abstract}
\bigskip
\date{\today}

\pacs{21.10.-k, 21.60.-n, 25.40.Hs, 21.10.Ky}

\maketitle

\section{Introduction}

The nucleus  $^{228}$Th  is located in a region where strong octupole correlations are important
in the properties already of the low-lying excitations. Besides the interplay of collective and
single-particle excitations, which takes place in deformed rare earth nuclei, the reflection
asymmetry additionally complicates the picture of excitations. Already in an earlier publication
\cite{Mah72}, a conclusion was made that the nature of the first excited 0$^+$ states in the
actinide nuclei is different from that in the rare earth region, where they are due to the
quadrupole vibration. The strong excitations in the (p,t)-reaction suggest that these states
represent a collective excitation different from the $\beta$-vibration. Decay modes of the
levels of the band on the first excited 0$^+$ state in $^{228}$Th have led to the suggestion
that this band might predominantly have an octupole two-phonon structure \cite{Dal87}. One has
to expect a complicated picture at higher excitations: residual interactions could mix the
one-phonon and multiphonon vibrations of quadrupole and octupole character with each other and
with quasiparticle excitations. Detailed experimental information on the properties of such
excitations is needed. On the experimental side, two-neutron transfer reactions are very useful.
On the theoretical side, a test of the advanced interacting boson model (IBM) and  a microscopic
approach, such as the quasiparticle-phonon model (QPM), would be very interesting.

After the first observation of a large number of  excitations  with the $L=0^+$ transfer in the
(p,t) reaction seen in the odd nucleus $^{229}$Pa \cite{Lev94}, it was logical to investigate
such excitations in the even-even nucleus $^{228}$Th, since $^{229}$Pa corresponds to $^{228}$Th
+ p, as well as in other actinide nuclei. Such measurements were carried out for the nuclei
$^{228}$Th, $^{230}$Th and $^{232}$U, and the results of a limited analysis have been published
in \cite{Wir04} (besides the earlier preliminary study of $^{228}$Th and $^{232,234,236}$U in
\cite{Bal96}). The paper \cite{Wir04}  concentrated only on the energies of the excited $0^+$
states in these actinide nuclei and  the (p,t) transfer strengths to these states.
 The (p,t) reaction, however, gives much more extensive information on  specific excitations
 in these nuclei, which was not analyzed previously. Such information was obtained for
$^{230}$Th in our paper \cite{Lev09} after detailed analysis of the experimental data
from  the (p,t) reaction. For the $0^+$ excitation, we were able to derive additional
information on the moments of inertia, which can be useful in clarifying the structure of these
excitations.
 In this paper we present  the results of a careful and detailed analysis of the experimental
 data from the high-resolution study of the $^{230}$Th(p,t)$^{228}$Th reaction carried out
 to obtain deeper insight into all excitations in
$^{228}$Th.  The total picture for $^{228}$Th has to differ from the one for $^{230}$Th, since
the first one is considered as an octupole soft and the latter as a vibration-like nucleus. It
would be interesting to compare the $0^+$ excitations in the even nucleus $^{228}$Th and the odd
nucleus $^{229}$Pa,  the data for which in the low-energy part of excitations are known from the
publication \cite{Lev94}.

\begin{figure*}
\begin{center}
\epsfig{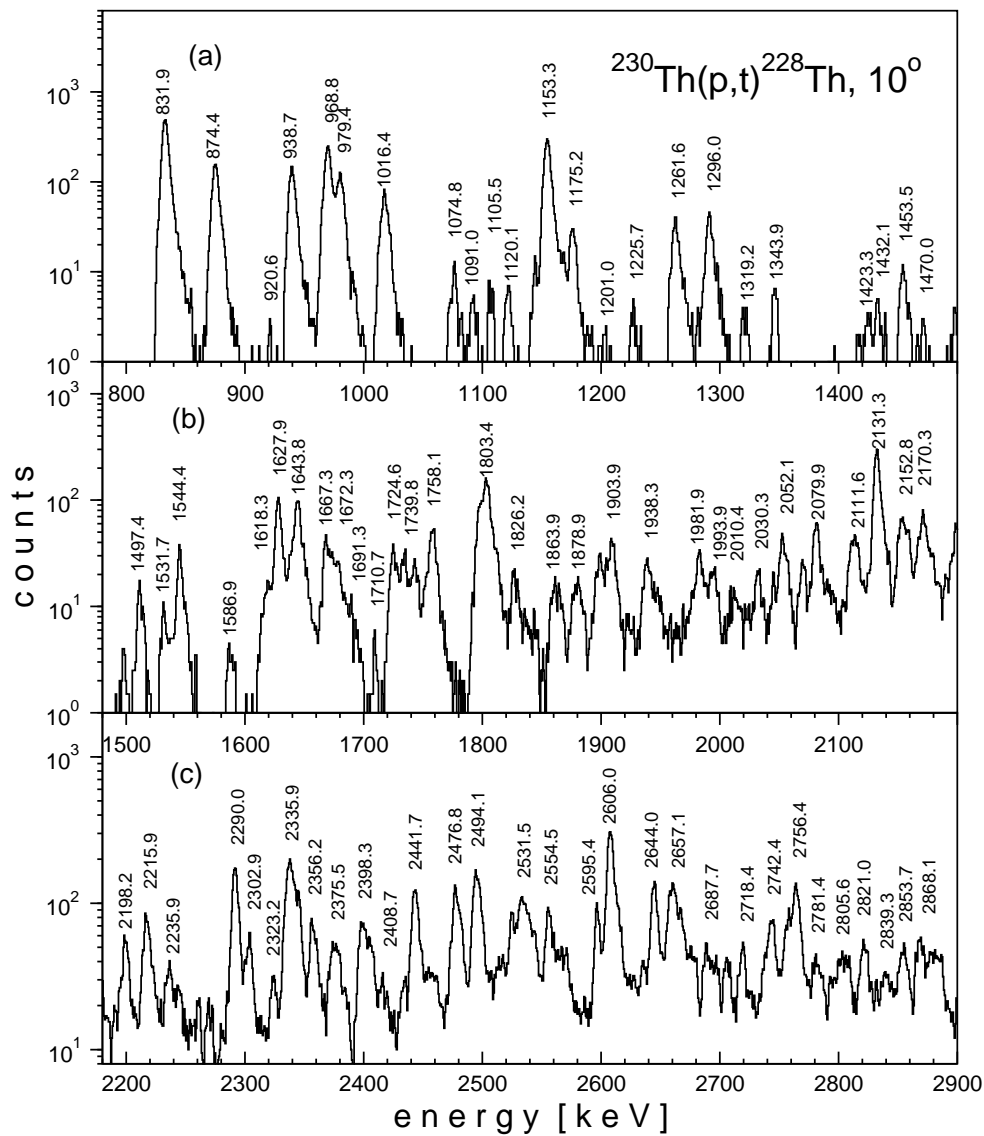} \caption{\label{fig:specTh228_10deg} Triton energy
spectrum from the $^{230}$Th(p,t)$^{228}$Th reaction (E$_p$=25 MeV) in
 logarithmic scale for a detection angle of 10$^\circ$. Some strong lines
 are labelled with their   corresponding level energies in keV.}
\end{center}
\end{figure*}

Information on excited states of $^{228}$Th prior to this study was obtained mainly from the
$\alpha$-decay of $^{232}$U, the $\beta$- and EC-decay of $^{228}$Ac and $^{228}$Pa, as well as
from the ($\alpha$, xn$\gamma$)-reaction. The most complete information was obtained from the
$\beta$-decay of $^{228}$Ac reported by Dalmasso et al. \cite{Dal87} and from the EC-decay study
of $^{228}$Pa by Weber et al. \cite{Web98}. The lowest collective bands in $^{228}$Th were
studied in the ($\alpha$, xn$\gamma$)-reaction \cite{Schu86}.  A total of 58 levels were
reported in \cite{Dal87} and 80 levels were observed in \cite{Web98} below 2.1 MeV, connected by
more than 240 $\gamma$-rays that were established in these studies.

Present results, derived from the $^{230}$Th(p,t)$^{228}$Th reaction, lead to about 163 levels
in the energy range up to 3.25 MeV. Unfortunately, during the experiment the radioactive target
was destroyed and assignments were made only for 106 levels in the range up to 2.5 MeV. Energies
and cross sections for one angle were obtained additionally for 57 levels. Besides 0$^+$
excitations, where the number of reliable assignments could be increased for five states in
comparison with the preliminary analysis in publication \cite{Wir04}, information on the spins
for many other states was obtained. This information was essentially complementary to what was
known from publications \cite{Dal87, Web98}. Some levels are grouped into rotational bands, thus
allowing to derive the moment of inertia for some $0^+$, $2^+$ and $0^-$, $1^-$, $2^-$, $3^-$
bands. One of the results of Ref. \cite{Web98} was the establishment of the one-phonon
octupole-quadruplet with $K^{\pi} = 0^-, 1^-, 2^-, 3^-$ states. In this paper we suggested the
two-phonon octupole-quadruplet with $K^{\pi} = 0^+, 2^+, 4^+, 6^+$ states.

\section{\label{sec:ExAnRe} Experiment, analysis and experimental  results}

\subsection{ \label{sec:det_exp} Details of the experiment}

A radioactive target of  100 $\mu$g/cm$^2$ $^{230}$Th with half-life T$_{1/2}$ = 8$\cdot10^4$
years, evaporated on a 22
 $\mu$g/cm$^2$ thick carbon backing, was bombarded with 25 MeV protons
at an intensity of 1-2 $\mu$A from the Tandem accelerator of the Maier-Leibnitz-Labor of the
Ludwig-Maximilians-Universit\"at and Technische Universit\"at M\"unchen. The isotopic purity of
the target was about 99\,\%. The tritons were analyzed with the Q3D magnetic spectrograph  and
then
 detected in a focal plane detector. The focal plane detector is a multiwire
 proportional chamber with readout of a cathode foil structure for
 position determination and dE/E particle identification
 \cite{Zan91,Wir01}. The acceptance of the spectrograph was 11 msr,
except for the most forward angle of 5$^\circ$ with an acceptance of 6 msr. The resulting triton
spectra have a resolution of
 4--7 keV (FWHM) and are background-free.
 The angular distributions of the cross sections were obtained from the
 triton spectra at ten laboratory angles from 5$^\circ$ to 40$^\circ$ for energies
 up to 1800 keV, but only at five angles from 7.5$^\circ$ to 30$^\circ$ for energies
 from 1800 to 2500 keV. The energies and cross sections for the states from 2500 keV
 to 3250 keV were measured only for 10$^\circ$.

 A triton energy
spectrum measured at a detection angle of 10$^\circ$ is shown in
 Fig.~\ref{fig:specTh228_10deg}. The analysis of
 the triton spectra was performed with the program GASPAN \cite{Rie91}.
Measurements were carried out with two magnetic settings: one for
 excitations up to 1.75 MeV, and another one for the energy region
 from 1.6 MeV to 3.3 MeV, respectively.   For the calibration of the energy scale,
the triton spectra from the reactions $^{184}$W(p,t)$^{182}$W,
 $^{186}$W(p,t)$^{184}$W  and $^{234}$U(p,t)$^{232}$U were
 measured at the same magnetic settings. The levels
in $^{230}$Th known from the study \cite{Lev09} were also included in the calibration.

From 106 levels identified in the spectra, 60 levels were identified for all ten angles and 46
levels only for 5 angles. They are listed in Table~\ref{tab:expEI}. The energies and spins of
the levels as derived from this study are compared to known energies and spins, mainly from the
published data \cite{Dal87,Web98}. They are given in the first four columns. The ratios of cross
sections at angles 7.5$^\circ$ and 26$^{\circ}$ to the one at 16$^{\circ}$, given in the next
two columns, help to highlight the $0^+$ excitations (large values). The column labelled
$\sigma_{\mbox{integ.}}$ gives the cross section integrated in the region from 7.5$^{\circ}$ to
30$^{\circ}$, where the cross sections are measured for all level energies. The column titled
{$\sigma_{\mbox{exp.}}/\sigma_{\mbox{calc.}}$} gives the ratio of the integrated cross sections,
from experimental values and calculations in the DWBA approximation (see Sec. \ref{sec:DWBA}).
The last column lists the notations of the schemes used in the DWBA calculations: sw.jj means
one-step direct transfer of the $(j)^2$ neutrons in the (p,t) reaction; notations of the
multi-step transfers used in the DWBA calculations are displayed in Fig.~\ref{fig:schemes}.
Additionally, energies of 57 levels seen only in the spectrum measured at 10$^\circ$
 and corresponding cross sections are listed in Table~\ref{tab:expEsigma}.\\

\newcolumntype{d}{D{.}{.}{3}}
\renewcommand{\thefootnote}{\fnsymbol{footnote}}
\begin{longtable*}{ll l ccc cc r c c}
\caption{\label{tab:expEI} \normalsize Energies of  levels in $^{228}$Th, the level spin
assignments from the CHUCK analysis, the (p,t) cross sections integrated over the measured
values and the reference to the schemes used in the DWBA calculations
(see text for more detailed explanations).}\\
\hline\hline
\smallskip\\
\multicolumn{3}{c}{Level energy [keV]}&\multicolumn{2}{c}{$I^\pi$}&&\multicolumn{2}{c} {Cross
section ratios}& \enspace {$\sigma_{\mbox{integ.}}$}&{Ratio}&Way of
\smallskip\\
\multicolumn{2}{l}{This work}&\enspace[2,7]&[2,7] &This work&&{(7.5$^o$/16$^o$)}
&{(26$^o$/16$^o$)}&[{$\mu$b}]&
{$\sigma_{\mbox{expt.}}/\sigma_{\mbox{calc.}}$}&fitting\\
\smallskip\\
\hline
\endfirsthead
\caption{Continuation}\label{tab:expEI}\\
\hline\hline
\smallskip\\
\multicolumn{3}{c}{Level energy
 [keV]}&\multicolumn{2}{c}{$I^\pi$}&&\multicolumn{2}{c}
{Cross section ratios}& \enspace {$\sigma_{\mbox{integ.}}$}&{Ratio}&Way of
\smallskip\\
\multicolumn{2}{l}{This work}&\enspace[2,7]&[2,7]& This work&&{(7.5$^o$/16$^o$)}
&{(26$^o$/16$^o$)}&[{$\mu$b}]&
{$\sigma_{\mbox{expt.}}/\sigma_{\mbox{calc.}}$}&fitting\\
\smallskip\\
\hline
\endhead
\hline
\endfoot
\endlastfoot
~~~0.0 \it2  &&    ~~~0.00    & $0^+$         & $0^+$         && 5.83 & 5.61 & 165.56 &6.20& sw.gg  \\
~~57.8 \it2  &&    ~~57.76    & $2^+$         & $2^+$         && 1.59 &0.68 &37.07  &8.30&m1a.gg\\
~186.8 \it2  &&    ~186.83    & $4^+$         & $4^+$         && 0.74 &0.38 &9.07   &1.90&m1a.gg\\
~328.0 \it2  &&    ~328.00    & $1^-$         & $1^-$         && 0.45 &0.66 &0.82   &0.50&m2a.gg\\
~378.2 \it2  &&    ~378.18    & $6^+$         & $6^+$         && 0.58 &0.71 &4.48   &1.60&m2a.gg\\
~396.9 \it2  &&    ~396.08    & $3^-$         & $3^-$         && 0.54 &0.33 &2.89   &0.56&m3a.gg\\
~519.2 \it3  &&    ~519.20    & $5^-$         & ($5^-$)              && 1.23 & 1.33 & 0.43   &0.90&sw.gg\\
~622.5 \it4  &&    ~622.50    & $8^+$         & $(8^+)$       && 0.20 & 0.94 & 0.26   &&\\
~695.6 \it3  &&    ~695.50    & $7^-$         & $(7^-)$       && 0.15 & 0.41 & 0.37   &&\\
~831.9 \it2  &&    ~831.83    & $0^+$         & $0^+$         && 12.06& 7.50 & 39.10  &360 &sw.ii\\
~874.4 \it2  &&    ~874.42    & $2^+$         & $2^+$         && 1.22 & 0.58 & 9.57   &160 &m1a.ii\\
~911.6 \it5  &&    ~911.80    & $10^+$        &               &&  &  & &  &\\
~920.6 \it5  &&    ~920.80    & $9^-$         &               &&  &  & &  &\\
~938.7 \it2  &&    ~938.55    & $0^+$         & $0^+$         && 18.38 &7.21 & 6.83   &8.20 &sw.ii\\
~943.8 \it4  &&    ~944.19    & $1^-$         & $1^-$         && 0.12 & 0.67 & 0.37   &1.00 &sw.gg\\
             &&    ~968.33    & $2^-$         &               &&  &  & &  &\\
             &&    ~968.43    & $4^+$         &               &&  &  & &  &\\
~968.8 \it2  &&    ~968.97    & $2^+$         & $2^+$         && 0.67 & 0.47 & 20.0   &132&sw.ig\\
~979.4 \it2  &&    ~979.50    & $2^+$         & $2^+$         && 0.78 & 0.59 & 9.25   &55.6&sw.ig\\
1016.4 \it2  &&    1016.43    & $2^+,3^-$     & $3^-$         && 0.80 & 0.47 & 5.37   &1.10&m3a.gg\\
             &&    1022.53    & $3^+$         &               &&  &  & &  &\\
             &&    1059.93    & $(3^-,4^+)4^-$ &$4^-$        &&  &  & &  &\\
1074.8 \it3  &&    1074.8    & $4^+$         & $4^+$          && 0.74 & 1.32 & 1.62   &0.26 &m1a.gg\\
1091.0 \it3  &&    1091.01    & $4^+$         & $4^+$         && 0.74 & 0.44 & 0.42   &0.10 &m1a.gg\\
1105.5 \it3  &&               &               & $6^+$         && 0.61 & 0.56 & 0.77   &21.0 &sw.ii\\
1120.1 \it3  &&    1120.1    & $0^+$         & $0^+$          && 2.63 & 3.71 & 1.24   &0.04 &sw.gg\\
             &&    1122.95    & $2^-$         &               &&  &  & &  &\\
1142.8 \it3  &&    1143.2    & $5^-$         & $5^-$          && 0.80 & 0.98 & 1.10   &26.0 &sw.jj\\
1153.3 \it3  &&    1153.48    & $2^+$         & $2^+$         && 0.65 & 0.49 & 23.89  &140  &sw.ig\\
1168.0 \it4  &&    1168.37    & $3^-$         & $3^-$         && 0.36 & 0.58 & 0.68   &1.00 &w.gg\\
             &&    1174.52    & $5^+$         &               &&  &  & &  &\\
1175.2 \it4  &&    1175.40    & $2^+$         & $2^+$         && 1.05 & 0.91 & 2.09   &13.0  &sw.ig\\
1201.0 \it9  &&    1200.54    & $3^+$         & $3^+$         && 0.31 & 1.10 & 0.40   &0.56  &m2a.gg\\
1225.7 \it6  &&               &               & $4^+$         && 1.00 & 0.64 & 0.25   &1.75  &sw.jj\\
             &&    1226.55    & $4^-$        &                &&  &  & &  &\\
1261.6 \it3  &&    1261.5    & $4^+$         & $4^+$          && 1.33 & 1.12 & 3.64   &67.0  &sw.ii\\
1270.2 \it6  &&    1270.0    &               & $6^+$          && 0.40 & 0.97 & 0.31   &0.15  &sw.gg\\
1290.4 \it3  &&    1290.2    & $4^+$         & $4^+$          && 1.14 & 0.88 & 3.59   &67.0  &sw.ii\\
1296.0 \it5  &&    1297.34    & $5^-$       & $(5^-)$       && 1.33 & 1.23 & 0.50   &1.00  &sw.gg\\
1319.2 \it4  &&               &               & $(2^+)$       && 0.74 & 1.08 & 0.24   &1.50  &sw.ig\\
1343.9 \it5  &&    1344.03    & $3^-$         & $3^-$         && 0.77 & 0.33 & 0.31   &0.08  &m3a.gg\\
             &&    1393.4    & $1^+,2,3^-$    & $(1^+)$               &&  &  & &  &\\
1415.8 \it6  &&    1415.92    & $2^+,3^-$     &$(3^-)$        && 1.15 & 1.20 & 0.05   &2.80  &sw.jj\\
1423.8 \it5  &&               &               & $(2^+)$       && 2.20 & 1.33 & 0.16   &0.03  &m1a.gg\\
1432.1 \it5  &&    1431.98    & $3^+,4^+$         & $4^+$         && 1.61 & 1.17 & 0.21   &6.80  &sw.ii\\
             &&    1448.80    & $3,4^-$       &               &&  &  & &  &\\
             &&    1450.29    & $3^-,4^-$     &               &&  &  & &  &\\
1453.5 \it5  &&               &               & $(3^-)$       && 0.61 & 0.63 & 1.34   &1.80  &sw.jj\\
1470.0 \it5  &&               &               & $(6^+)$       && 0.94 & 1.81 & 0.19   &0.01  &m3a.gg\\
1497.4 \it4  &&    1497.7    & $4^+,5^-$       & $(5^-)$     && 1.07 & 0.91 & 0.37   &0.56  &sw.gg\\
1511.2 \it3  &&               &               & $0^+$         && 7.96 & 6.96 & 2.13   &1.10 &sw.ig\\
1531.7 \it3  &&               &               & $0^+$         && 2.21 & 0.83 & 0.47 &2.60 & sw.ii\\
 \hspace{5mm}  plus   &&    1531.48    & $3^+$         &  $3^+$        &&      &      &        & 0.02    &m2a.gg\\
             &&    1539.13    & $(2,3)$         &             &&  &  & &  &\\
1544.4 \it3  &&               &               & $2^+$          && 1.27 & 0.65 & 1.61   &1.53  &m1a.gg\\
             &&    1581.0    & $(2^-)$        &               &&  &  & &  &\\
1586.9 \it4  &&              &                & $2^+$         && 0.98 & 0.71 & 0.31   &1.00  &sw.jj\\
             &&    1588.33    & $4^-$         &               &&      &      &        &  &\\
1613.0 \it5  &&               &               & $4^+$       && 1.06 & 1.26 & 0.54   &12.0  &sw.ii\\
1618.3 \it5  &&    1617.74    & ($3,4)^+$     & $4^+$         && 0.88 & 0.76 & 1.22   &0.16  &m2a.ii\\
1627.9 \it3  &&               &               & $0^+$         && 7.44 & 5.21 & 9.66   &10.0  &sw.ig\\
1638.4 \it4  &&    1638.25    & $2^+$         & $2^+$         && 0.59 & 0.37 & 1.45   &23.5  &sw.ii\\
             &&    1643.18    & $(2,3)^-$     &               &&  &  & &  &\\
1643.8 \it3  &&    1643.8    & $(2,3,4)^+$    & $4^+$         && 1.58 & 1.08 & 8.54   &160  &sw.ii\\
             &&    1645.89    & $3^+$          &               &&  &  & &  &\\
1651.4 \it3  &&               &               & $(3^-)$       && 0.08 & 0.79 & 0.86   &1.20  &sw.gg\\
1667.3 \it5  &&    1667.3     &               & $2^+$         && 0.71 & 0.61 & 3.17   &46.0  &sw.ii\\
1672.3 \it5  &&               &               & $2^+$         && 0.91 & 0.53 & 1.85   &3.80  &sw.jj\\
1678.4 \it5  &&    1678.4     &  $2,3,4^+$    & $2^+$       && 0.88 & 0.75 & 1.48   &19.5  &sw.ii\\
             &&   1682.70     & $(3,4)^+$     &               &&  &  & &  &\\
             &&   1683.74     & $(4^-)$       &               &&  &  & &  &\\
             &&   1688.39     & $3^+$         &               &&  &  & &  &\\
1691.3 \it4  &&               &               & $0^+$         && 2.66 & 2.06 & 1.26   &0.75  &sw.ig\\
             &&   1707.2     & $2,3^-$        &               &&  &  & &  &\\
1710.7 \it6  &&               &               & $0^+$         && 1.38 & 1.86 & 0.54   &0.02  &sw.gg\\
1724.6 \it4  &&   1724.29     & $2^+$         & $2^+$       && 1.06 & 0.66 & 2.73   &5.50  &sw.ii\\
1733.8 \it4  &&   1735.62     & $4^+$         & $4^+$       && 1.13 & 0.86 & 2.28   &3.50  &sw.ij\\
1742.8 \it4  &&   1743.86     & $3,4^+$         & $4^+$         && 0.81 & 0.56 & 1.36   &0.16  &m2a.gg\\
1750.7 \it3  &&               &               & $0^+$       && 1.27 & 1.84 & 1.75   &0.70  &sw.jj\\
1758.1 \it3  &&   1757.9      & $1^-,2, 3^-$  &  $2^+$       && 0.85 & 0.75 & 4.35   &26.0  &sw.ig\\
             &&   1758.24     & $(3,4)^+$     &              &&  &  & &  &\\
             &&   1760.25     & $(2,3,4)^+$   &              &&  &  & &  &\\
             &&   1795.9      &   $3^-,4^+$    &               &&  &  & &  &\\
1796.8 \it3  &&   1796.4      & $3^+,4,5^+$     & $4^+$       && 1.40 & 0.83 & 6.47   &89.0  &sw.ig\\
             &&   1797.65     & $(2^+,1^-)$   &               &&  &  & &  &\\
1803.0 \it4  &&   1802.9      &  $1^-,2,3^-$  & $2^+$         && 0.65 & 0.49 & 15.34  &90.0  &sw.ig\\
             &&   1804.60     & $(4^+)$       &               &&  &  & &  &\\
             &&   1811.5     & $1^-,2,3^-$   &               &&   & &  &  &\\
1812.7 \it4  &&               &               & $(6^+)$       && 1.35 & 1.63 & 0.62   &0.04  &sw.ig\\
             &&   1817.43     & $4^-$         &               &&  &  & &  &\\
             &&   1823.4     & $3^-,4,5$     &               &&  &  & &  &\\
1826.2 \it4  &&               &               & $(4^+)$       && 1.16 & 0.83 & 1.91   &7.50  &sw.ij\\
1840.3 \it8  &&   1842.2     & $2^+,3^-$     &                && 1.41 & 0.33 & 0.21   &  &\\
1858.6 \it5  &&               &               & $(6^+)$       && 0.65 & 1.19 & 1.28   &0.06  &sw.ig\\
1863.9 \it5  &&    1864.8    & $1^-,2,3^-$     & $(2^+)$       && 0.75 & 0.79 & 1.47   &8.10  &sw.ig\\
             &&   1876.5     & $3^-,4,5^-$   &               &&  &  & &  &\\
1878.9 \it5  &&   1879.0     & $3^-,4,5^-$     & $(3^-)$       && 1.05 & 0.91 & 1.93    &110  &sw.ii\\
             &&   1892.98     & $3^+$         &               &&  &  & &  &\\
1898.2 \it4  &&   1899.98     & $2^+$         & $(2^+)$       && 0.84 & 0.81 & 2.55   &140  &sw.ii\\
             &&   1901.90     & $4^+$         &               &&      &      &        &      &     \\
1903.9 \it4  &&               &               & ($6^+$)       && 0.69 & 1.58 & 1.54   &0.07  &sw.gg\\
             &&   1906.78     & $(2^+,1^-)$   &               &&  &  & &  &\\
             &&   1908.4     & $3^-$          &               &&  &  & &  &\\
1908.9 \it7  &&               &               & $0^+$         && 2.17 & 1.91 & 4.56   &1.30  &sw.jj\\
             &&   1924.1     & $2^-,3,4$          &               &&  &  & &  &\\
             &&   1924.6     & $(4,5^+)$          &               &&  &  & &  &\\
1925.4 \it4  &&   1925.20     & $4^+$          &  $4^+,5^-$    && 0.61 & 1.73 & 0.54   &21.0  &sw.ii\\
             &&   1928.54     & $3^+$         &               &&  &  & &  &\\
             &&   1937.16     & $(3,4)^+$   &               &&  &  & &  &\\
1938.3 \it4  &&   1938.9     & $4^+$        & ($4^+$)       && 1.06 & 0.76 & 1.99   &0.67  &m2a.gg\\
             &&   1944.85     & $3^+$         &               &&  &  & &  &\\
             &&   1945.8     & $4^+,5^-$      &               &&  &  & &  &\\
1947.8 \it7  &&               &               & ($2^+$)       && 1.02 & 0.75 & 0.77   &3.50  &sw.ig\\
             &&   1949.7     & $1^+,2,3^+$    &               &&  &  & &  &\\
1959.7 \it6  &&   1958.5      & $2^+$         & $(2^+)$       && 0.10 & 1.69 & 0.43   &1.50  &sw.ig\\
             &&   1964.90     & $(2^+)$       &               &&  &  & &  &\\
1971.7 \it4  &&               &               & $(2^+,3^-)$       && 0.66 & 0.81 & 0.79   &3.10  &sw.ig\\
             &&   1974.20     & $4^+$         &               &&  &  & &  &\\
1981.9 \it4  &&   1981.97     & $3,4^+$       &  $(3^-)$      && 1.68 & 0.77 & 1.70   & 2.60 &sw.gg\\
             &&   1987.46     & $4^+$         &               &&  &  & &  &\\
1993.9 \it5  &&               &               & ($3^-$)       && 0.97 & 0.72 & 1.80   &2.80  &sw.gg\\
2010.4 \it6  &&   2010.15     & $2^+,3,4^+$   & $(2^+)$       && 0.46 & 0.43 & 0.76   &13.0  &sw.ig\\
             &&   2013.6     & $(3,4)^+$     &               &&  &  & &  &\\
             &&   2016.75     & $4^+,5^-$     &               &&  &  & &  &\\
             &&   2022.73     & $2^+$         &               &&  &  & &  &\\
2030.3 \it4  &&   2029.6     & $(2^+)$        & $2^+$         && 0.54 & 0.23 & 0.84   &16.0  &sw.ig\\
             &&   2037.0     & $(3,4)^+$      &               &&  &  & &  &\\
2044.7 \it5  &&               &               & $0^+$         && 9.22 & 4.56 & 0.57   &3.10  &sw.ii\\
2052.1 \it4  &&               &               & ($6^+$)       && 0.72 & 1.30 & 3.70   &180  &sw.ii\\
2069.6 \it5  &&               &               & $2^+$         && 0.76 & 0.56 & 1.38   &6.10  &sw.ig\\
2079.9 \it5  &&               &               & $0^+$         && 17.08 & 13.13 & 4.62 &25.9  &sw.ii\\
2091.2 \it7  &&               &               & $(6^+)$       && 0.62 & 0.82 & 1.20   &35.0  &sw.ii\\
2111.6 \it5  &&               &               & $(2^+)$       && 0.70 & 0.71 & 2.57   &11.0  &sw.ig\\
             &&   2123.1     & $(2^+,3^-)$   &               &&  &  & &  &\\
2131.3 \it6  &&               &               & $0^+$         && 6.84 & 4.53 & 24.80  &120  &sw.ii\\
2152.8 \it4  &&               &               & $(4^+)$       && 1.30 & 0.90 & 4.13   &98.0  &sw.ii\\
2159.4 \it6  &&               &               & $0^+$       && 3.78 & 1.54 & 1.18   &8.10  &sw.ii\\
2170.3 \it4  &&               &               & $(2^+)$       && 1.00 & 0.85 & 5.61   &26.0  &sw.ig\\
2198,2 \it4  &&               &               & $2^+$         && 0.59 & 0.62 & 3.81   &19.5  &sw.ig\\
2215.9 \it4  &&               &               &  ($4^+$)      && 1.40 & 1.15 & 6.00   &130  &sw.ii\\
2235.2 \it7  &&               &               &  ($4^+$)      && 0.98 & 0.86 & 2.82   &61.0  &sw.ii\\
2290.0 \it7  &&               &               & $0^+$         && 9.96 & 5.75 & 11.00  &61.0  &sw.ii\\
2302.9 \it5  &&               &               &  $(4^+)$      && 1.09 & 0.84 & 2.75   &62.0  &sw.ii\\
2323.2 \it5  &&               &               & $2^+$         && 0.41 & 0.62 & 2.24   &16.0  &sw.ig\\
2335.9 \it5  &&               &               & ($4^+$)       && 2.13 & 1.65 & 17.10  &370  &sw.ig\\
             &&               &               & ($0^+$)       &&      &      &  4.50$^a$  & 25.0 &sw.gg+14\\
2344.2 \it5  &&               &               &  ($3^-$)     && 0.77 & 0.58 & 6.65   &10.0   &sw.gg\\
2356.2 \it5  &&               &               &  ($2^+$)      && 0.63 & 0.61 & 4.61   &21.5  &sw.ig\\
2375.5 \it8  &&               &               &  ($2^+$)      && 0.78 & 0.60 & 4.87   &22.0  &sw.ig\\
2398.3 \it9  &&               &               &  ($3^-$)      && 0.76 & 0.75 & 7.36   &11.0  &sw.gg\\
2408.8 \it9  &&               &               &  ($4^+$)      && 1.87 & 1.28 & 2.34   &60.0  &sw.ii\\
2441.7 \it5  &&               &               &  ($2^+$)      && 0.71 & 0.51 & 10.32  &47.0  &sw.ig\\
2456.8 \it5  &&               &               &  $0^+$        && 16.18 & 1.27 & 0.53  &5.20  &sw.ii\\
2476.7 \it5  &&               &               &  ($2^+$)      && 0.62 & 0.52 & 10.38  &48.0  &sw.ig\\
2494.1 \it5  &&               &               &  ($2^+$)      && 0.65 & 0.47 & 12.74  &63.5  &sw.ig\\
\hline \\
\end{longtable*}
\footnotesize \noindent
 $^a$ The value after subtracting a constant of 14 $\mu$b (see text in
Sec.~\ref{sec:DWBA} for explanation. \\
\normalsize

\newcolumntype{d}{D{.}{.}{2}}
\begin{table}
\caption
    {\label{tab:expEsigma} Energies and cross sections of the $^{230}$Th(p,t)$^{228}$Th
    reaction for the states for which measurements were carried out only at 10$^o$.}
 \begin{ruledtabular}
    \begin{tabular}{lclclc}
    \smallskip\\
E [keV] & $d\sigma/d\Omega$ & E [keV] & $d\sigma/d\Omega$ & E [keV] & $d\sigma/d\Omega$\\
\smallskip\\
\hline
\smallskip\\
2513.5 \it7 & 2.00 & 2742.3 \it4 &5.50 & 3035.6 \it9 & 0.95\\
2531.5 \it7 & 6.60 & 2763.7 \it4 &8.60 & 3046.4 \it6 & 2.10\\
2536.8 \it9 & 3.20 & 2781.4 \it5 &1.75 & 3059.2 \it5 & 2.15\\
2542.4 \it9 & 1.85 & 2798.6 \it8 &1.55 & 3075.2 \it5 & 2.20\\
2554.5 \it5 & 6.00 & 2805.6 \it7 &2.00 & 3085.2 \it8 & 1.25\\
2566.3 \it6 & 2.20 & 2821.0 \it5 &2.90 & 3097.0 \it6 & 3.10\\
2595.4 \it5 & 5.40 & 2839.3 \it6 &1.30 & 3104.7 \it6 & 3.40\\
2606.1 \it5 & 23.5 & 2853.7 \it5 &2.75 & 3112.7 \it11 &1.70\\
2615.1 \it9 & 0.15 & 2868.1 \it5 &3.20 & 3119.9 \it9 &2.30\\
2634.8 \it5 & 1.60 & 2877.5 \it8 &1.80 & 3128.2 \it10 &1.25\\
2644.0 \it3 & 9.20 & 2883.7 \it9 &1.60 & 3158.8 \it8 &1.50\\
2657.1 \it4 & 5.20 & 2918.8 \it6 &1.85 & 3165.7 \it6 &2.00\\
2660.1 \it5 & 6.00 & 2927.4 \it5 &3.25 & 3186.0 \it6 &2.00\\
2667.1 \it5 & 3.30 & 2936.8 \it9 &1.40 & 3195.2 \it6 &2.60\\
2676.0 \it6 & 67.2 & 2945.3 \it9 &1.35 & 3209.6 \it12&1.40\\
2688.4 \it4 & 2.10 & 2955.1 \it8 &1.25 & 3214.8 \it9 &2.20\\
2695.6 \it7 & 1.10 & 2993.1 \it12 &1.00 &3225.0 \it20 &0.50\\
2705.5 \it5 & 1.35 & 2999.5 \it10 &1.50 &3232.9 \it13 &1.20\\
2718.4 \it5 & 2.10 & 3014.3 \it11 &0.80 &3239.9 \it8  &3.40\\
   \end{tabular}
\end{ruledtabular}
\end{table}

\subsection{\label{sec:DWBA} DWBA analysis}

The spins of the excited states in the final nucleus $^{228}$Th were assigned via an analysis of
the angular distributions of tritons from the (p,t) reaction. The angular distributions for
$0^+$ excitations have a steeply rising cross section at very small reaction angles, and a sharp
minimum at a detection angle of about 14$^\circ$.  This pronounced feature helped to identify
most of these states in complicated and dense spectra, even without fitting experimental angular
distributions. No complication of the angular distributions was expected, since the excitation
of  $0^+$ states predominantly proceeds via a one-step process. This is not the case for the
excitation of states with other spins, where multi-step processes could play a very important
role.

\begin{figure*}
\begin{center}
\epsfig{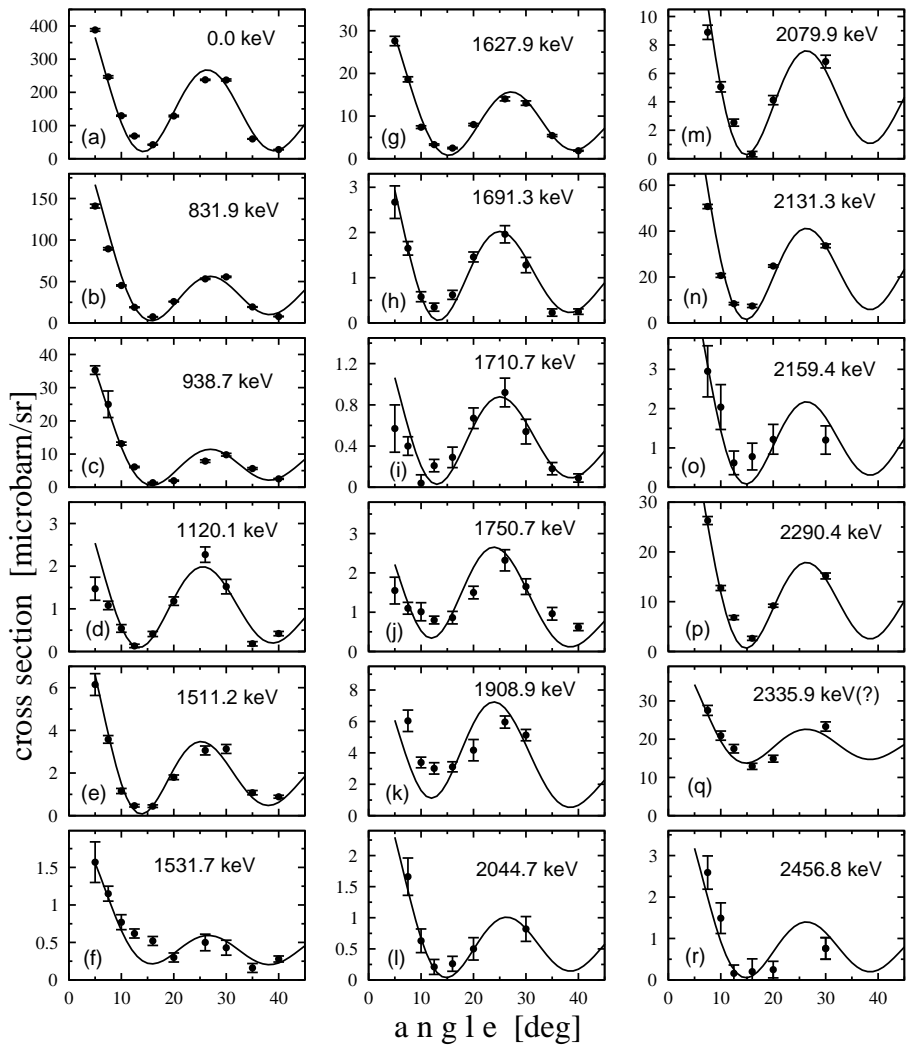}
    \caption{\label{fig:angl_distr_0+}
    Angular distributions of assigned $0^+$ states in $^{228}$Th and
     their fit with  CHUCK3 one-step calculations. The  $(ij)$ transfer
     configurations used in the calculations for the best fit are
    given in Table~\ref{tab:expEI}. See text for further information. }
\end{center}
\end{figure*}


The identification of these states is possible by fitting the experimental angular distributions
with those calculated in the distorted-wave Born approximation (DWBA). The potential parameters
suggested by Becchetti and Greenlees \cite{Bec69} for protons and by Flynn {\it et al.}
\cite{Fly69} for tritons were used in the calculations. These parameters have been tested via
their description of angular distributions for the ground states of $^{228}$Th, $^{230}$Th and
$^{232}$U  \cite{Wir04}.  Minor changes of the parameters for tritons were needed only for some
$3^-$ states, particularly for the states at 396.9 and 1016.4 keV. For these states, the triton
potential parameters suggested by Becchetti and Greenlees \cite{Bec71} were used.
 For each state the binding energies of the two neutrons are calculated to match the outgoing
 triton energies. The corrections to the reaction energy are introduced depending
 on the excitation energy. For more details see \cite{Lev09}.

A problem arising in such calculations is the lack of prior knowledge of the microscopic
structure of these states. We can assume, however, that  the overall shape of the angular
distribution of the cross section is rather independent of the specific structure of the
individual states, since the wave function of the outgoing tritons is restricted to the nuclear
exterior and therefore to the tails of the triton form factors. To verify this assumption, DWBA
calculations of angular distributions for different $(j)^2$ transfer configurations to states
with different spins were carried out in our previous paper \cite{Lev09}. The coupled-channel
approximation (CHUCK3 code of Kunz \cite{Kun}) was used in these calculations. Indeed, the
calculated angular distributions are very similar. Nevertheless, they depend
 to some degree on the transfer configuration, the most pronounced being found for
 the 0$^+$ states, what is confirmed  by the experimental angular distributions.
 The best reproduction of the angular distribution for the ground state was obtained for
 the transfer of the $(2g_{9/2})^2$ configuration in the one-step process.
 This orbital is close to the  Fermi surface and was considered as
the most probable one in the transfer process.  Other transfer
 configurations that might be of importance are $(1i_{11/2})^2$ and
 $(1j_{15/2})^2$, also near the Fermi surface. Better reproduction of
the angular distribution for some $0^+$ states is obtained just for these configurations. The
main features of the angular distribution shapes for 2$^+$ and 4$^+$ states are even more weakly
dependent on the transfer configurations. Nevertheless the $(2g_{9/2})^2$, $(1i_{11/2})^2$ and
$(1j_{15/2})^2$ configuration, alone or in combination, were used in the calculations for these
states too.

Results of  fitting the angular distributions for the states assigned as $0^+$ excitations are
shown in Fig.~\ref{fig:angl_distr_0+}.  The agreement between the fit and the data is
excellent for most of the levels. Remarks are needed only for the levels at 1531.7 and 2335.9
keV. The spin $3^+$ was assigned to the level at 1531.47 keV in \cite{Dal87,Web98}. A level at
the close-lying energy of 1531.7 keV has been observed also in the (p,t) reaction, but the
angular distribution of tritons cannot be fitted  by calculations for transition to the $3^+$
state. The maximum cross section for forward angles suggests the presence of a $0^+$ excitation,
though the angular distribution  fitted by a calculation for transfer of the $(1i_{11/2})^2$
configuration to the $0^+$ state is not perfect. A satisfactory fit of the experimental angular
distribution was obtained assuming overlapping states with spins $0^+$ and $3^+$
(Fig.~\ref{fig:angl_distr_0+}), thus confirming the assignment for both states. An ambiguous
picture is observed for the 2335.9 keV state, where the angular distribution is measured for a
limited range of angles.  The fitting agreement is perfect for a transition to the 4$^+$ state,
but the cross section is surprisingly large for a 4$^+$ state, twice larger than for the $4^+$
member of the g.s. band. Therefore,  the possibility of a 0$^+$ excitation can not be excluded,
but the experimental angular distribution is fitted for a $0^+$ state only with adding a
constant  of 14 $\mu$b. This ambiguity can be resolved by measurements in wider angular regions.

\begin{figure}
\begin{center}
\epsfig{file=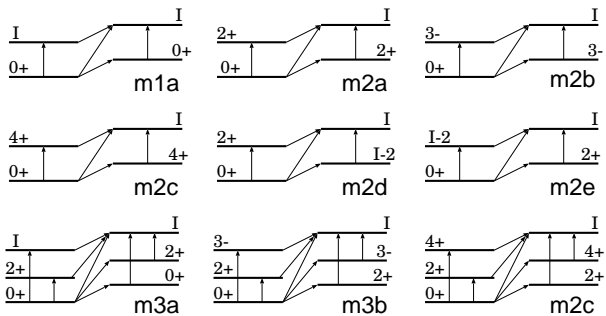, width=8cm,angle=0}
    \caption{\label{fig:schemes} Schemes of the CHUCK3 multi-step
    calculations tested with spin assignments of  excited states
    in $^{230}$Th (see Table~\ref{tab:expEI}).}
\end{center}
\end{figure}

\begin{figure*}
\begin{center}
\epsfig{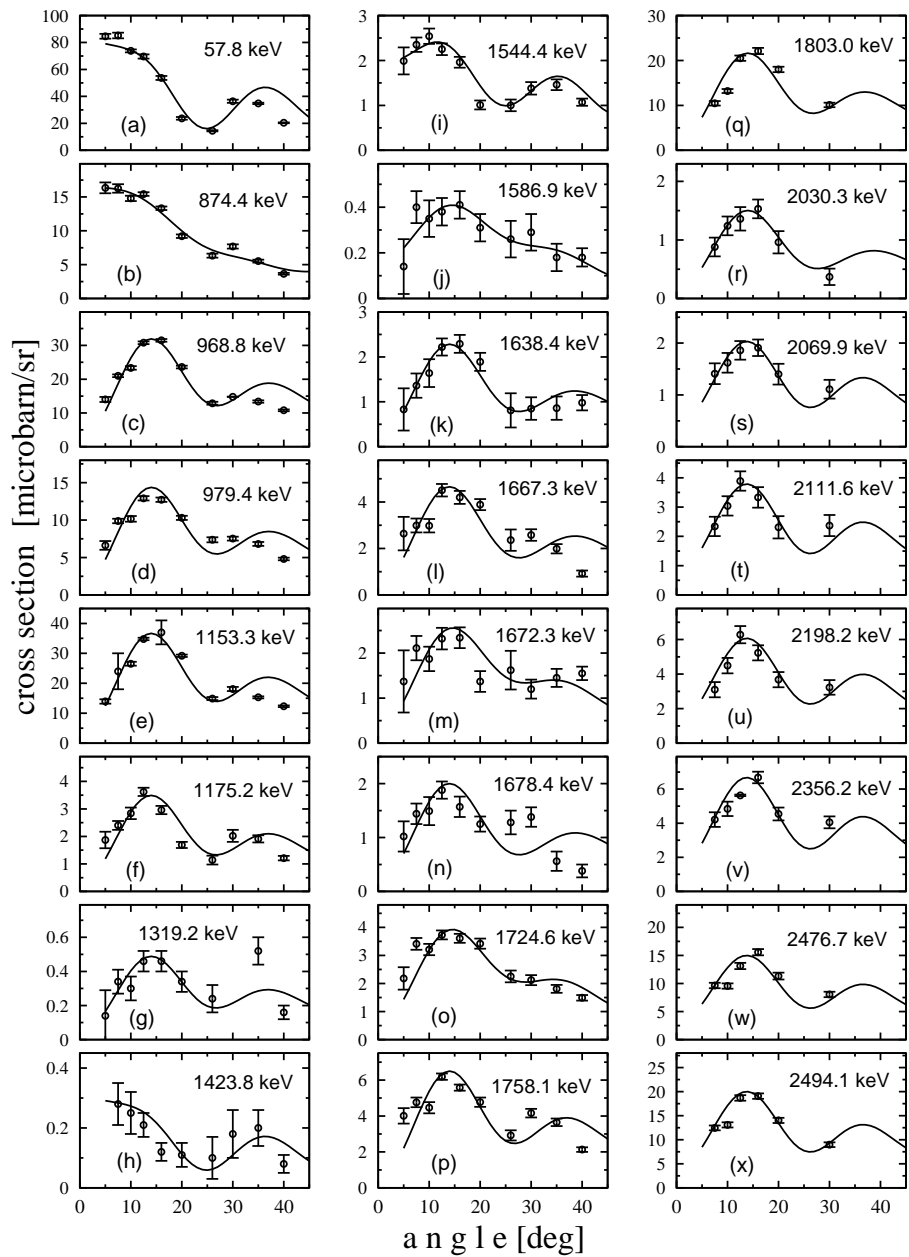}
    \caption{\label{fig:angl_distr_natur}Angular distributions of assigned $2^+$ states in $^{228}$Th and
     their fit with  CHUCK3  calculations.
     The  $(ij)$ transfer configurations and schemes used
    in the calculations for the best fit are given in Table~\ref{tab:expEI}.}
\end{center}
\end{figure*}

Thus we can make firm $0^+$ assignments for 17 states for energies excitations below 2.5 MeV,
 in comparison with 12 states found in the preliminary analysis of the
 experimental data \cite{Wir04}. Of course, some higher lying 0$^+$ levels are lost because of the
 cutoff of the investigated energy region.
But as follows from a similar study  for $^{230}$Th, only a few $0^+$ states are observed above
 2500 keV, where the density of $0^+$ excitations decreases for higher energies
 (or else that the cross section of such excitations is very low and they are hidden
 in very dense and complicated spectra). Therefore, we can compare 24 $0^+$ states in $^{230}$Th
 with only 17 $0^+$ states in $^{228}$Th in the same energy region.

\begin{figure*}
\begin{center}
 \epsfig{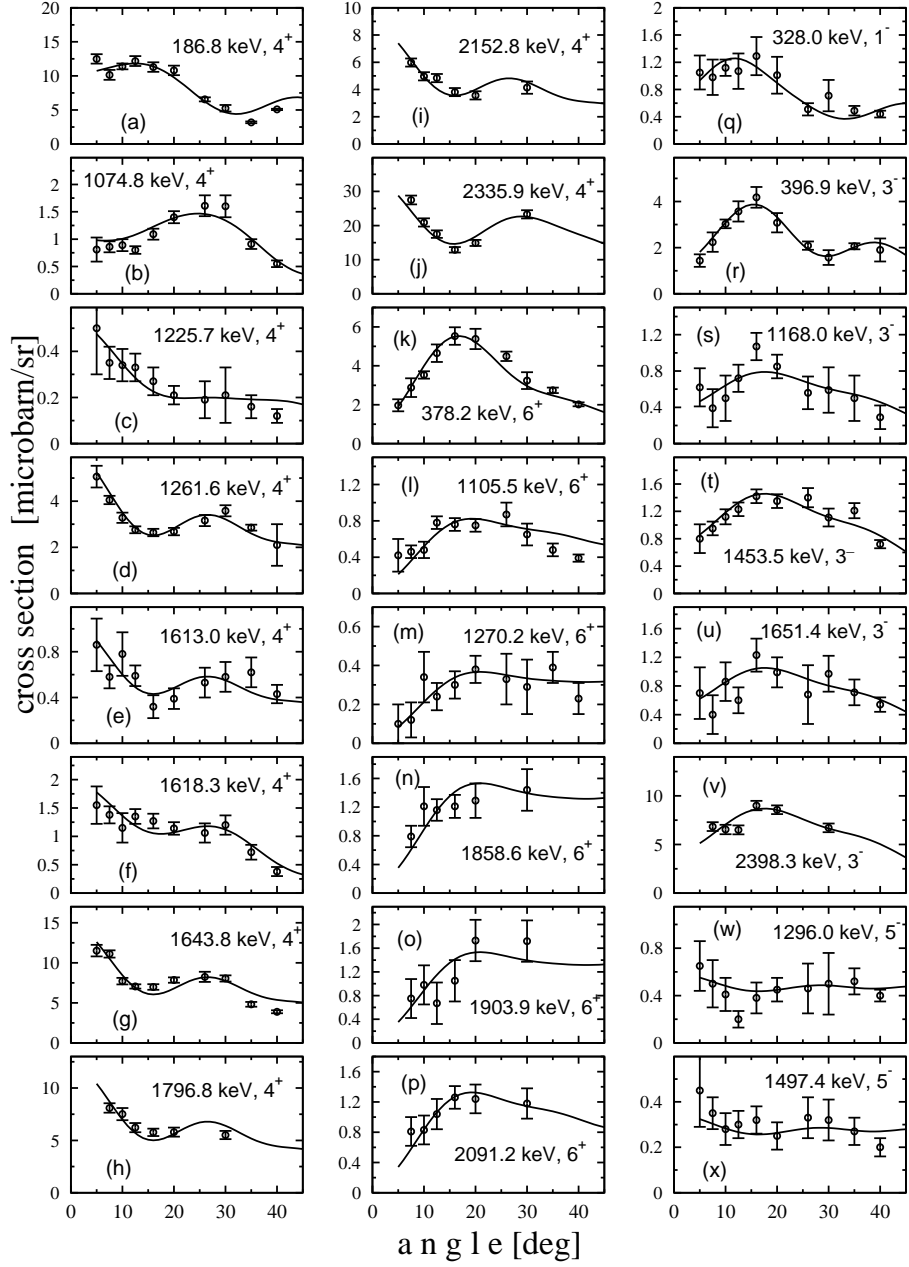}
    \caption{\label{fig:angl_distr_unnatur}Angular distributions
    of some assigned states in $^{228}$Th and their fit with
    CHUCK3 calculations: $4^+$ and $6^+$ with positive parity
    and $1^-$, $2^-$ and $3^-$ with negative parity.
    The  $(ij)$ transfer configurations and schemes used
    in the calculations for the best fit are given in Table~\ref{tab:expEI}.}
\end{center}
\end{figure*}

Similar to  $0^+$ excitations,   the one-step transfer calculations give a satisfactory fit of
angular distributions for about 80\% of the states with spins different from  $0^+$,  but about
20\% of these states need the inclusion of multi-step excitations. Multi-step excitations have
to be included to fit the angular distributions already for the $2^+$, $4^+$ and $6^+$ states of
the g.s. band. Fig.~\ref{fig:schemes} shows the schemes of the multi-step excitations, tested
for every state in those cases, where one-step transfer did not provide a successful fit.
Fig.~\ref{fig:angl_distr_natur} demonstrates the quality of the fit of some  different-shaped
angular distributions for the excitation of states with  spin $2^+$ by calculations assuming
one-step and one-step plus two-step excitations, respectively. Results of similar fits for the
states assigned as $4^+$, $6^+$ and $1^-$, $3^-$, $5^-$ excitations are shown in
Fig.~\ref{fig:angl_distr_unnatur}.

\begin{figure*}
\begin{center}
\epsfig{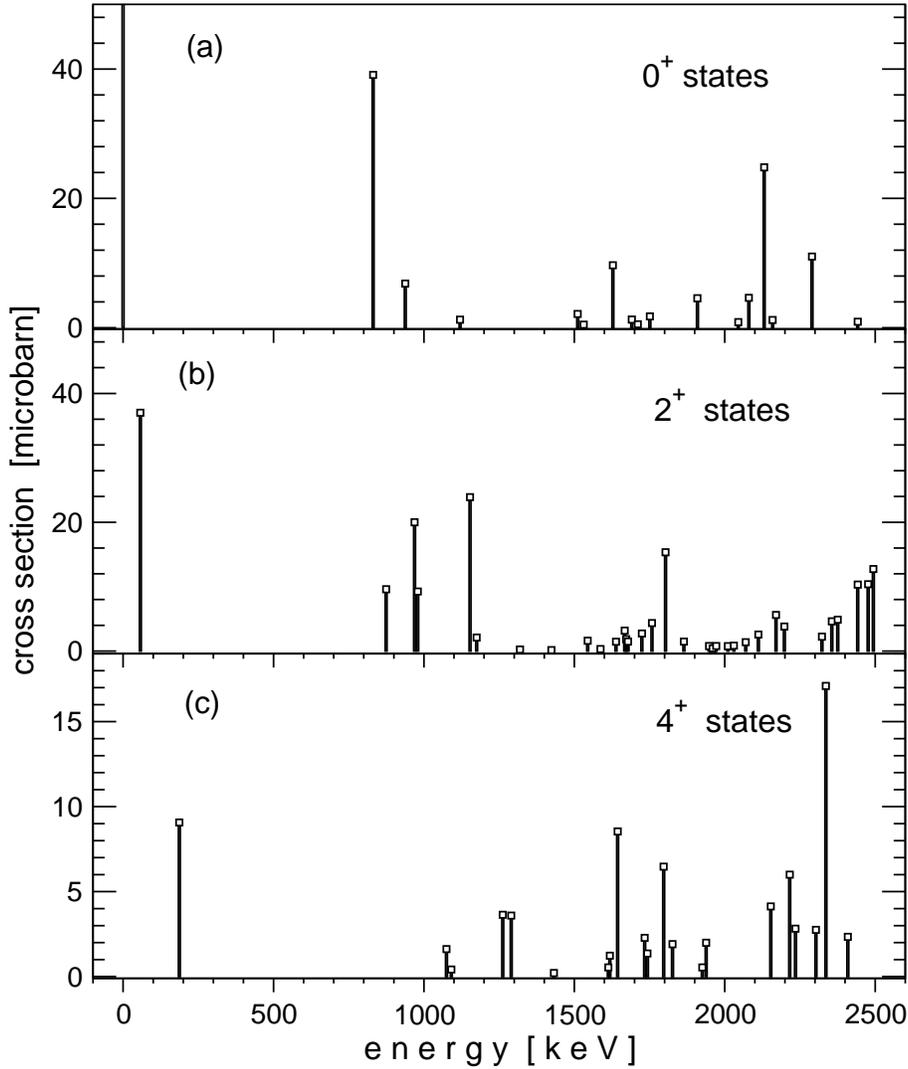}
    \caption{\label{fig:strength} Experimental  distribution of
    the (p,t) strength integrated in the angle region 7.5$^\circ$  - 30$^\circ$
    for 0$^+$, 2$^+$ and 4$^+$ states in $^{228}$Th.}
\end{center}
\end{figure*}

The assignments of the spins resulting from such fits are presented in Table~\ref{tab:expEI},
together with other experimental data.  Special comments are needed for the column displaying the ratio
$\sigma_{exp}/\sigma_{cal}$. Calculated cross sections for the specific transfer configurations
differ very strongly.  Since we have no a priori knowledge of the microscopic structure of
the excited states, and thus do not know the relative contributions of the specific $(j)^2$
transfer configurations to each of these states, these ratios cannot be considered as
spectroscopic factors. Nevertheless, a very large ratio, such as in the case of the
$(1i_{11/2})^2$ transfer configurations used in the calculation for some $0^+$ and even $2^+$
and $4^+$ states, is unexpected. Surprisingly, the shape just for this neutron configuration
gives the best agreement with  experiment.

Some additional comments on Table~\ref{tab:expEI} are needed.
In all cases, where the firm assignment were known from the
previous studies  \cite{Dal87,Web98}, they are confirmed by the (p,t) angular distribution
analysis. In those cases, where two or three possible spin assignments were proposed earlier,
the (p,t) angular distribution analysis allows to select only one assignment almost in all
cases. For the energies above 2030 keV only the assignments from the (p,t) reaction are possible
at present. The following remarks are needed in those cases, where the assignments from
different publications are in contradiction, which can be removed using the data from the (p,t)
reaction.

\emph{938.7 keV.} Spin $2^-$ was assigned for this level in \cite{Dal87}. Our fit of the angular
distribution gives reliably spin $0^+$ in agreement with \cite{Web98}.

\emph{943.8 keV.} Spin $2^+$ was assigned for this level in \cite{Dal87}. The angular
distribution
 rejects this value and agrees with the assignment of spin $1^-$ accepted in  \cite{Web98}.

\emph{968-969 keV.}
 Three levels around 968 keV with spins  $2^+, 4^+, 2^-$ were identified in
\cite{Web98} and two levels with spins $2^+, 3^-$  in \cite{Dal87}. There is a discrepancy in
assignment for the 968.33 keV level as $3^-$ in \cite{Dal87} and as $2^-$ in \cite{Web98}. This
line is masked in the (p,t) spectrum by a strong line from the transition to the 969 keV level.
But since the angular distribution is very well described by a calculation leading to a $2^+$
state and is not disturbed by a transition to spin $3^-$, the assignment $2^-$ is preferable
(transition is weak).

\emph{1016.4 keV.}   The discrepancy in assignment for the 1016.4 keV level as $2^+$ in
\cite{Dal87} and as $3^-$ in \cite{Web98}
 can not be removed by the (p,t) angular distribution, since it can be fitted by a transition
 sw.ig to $2^+$ and m3a.gg to  $3^-$, respectively. However, transitions seen in the decay of $^{228}$Pa \cite{Web98}
 from this state to the $5^-$ and $4^+$ states leads to the assignment of $3^-$.
 We accepted this spin also due to strong arguments in \cite{Web98}, including the assignment of
spins $2^-$ and $1^-$ to the levels at 968.3 keV and 943.8 keV as members of the $K^{\pi}=1^-$
band.

\emph{1059.9 keV.}  From the tentative
 assignments $(4^+,3^-)$ in \cite{Dal87} and the firm assignment $4^-$ in \cite{Web98} for this level,
 the latter has to be additionally supported by the fact,
 that  the corresponding line in the (p,t) spectrum was not seen (transition
 to the state of unnatural parity).

\emph{1225.7 and 1226.56 keV.} Spin $4^-$ was assigned for the level 1226.56 keV in both
\cite{Dal87} and \cite{Web98}. The level with the close energy 1225.7 keV is seen in the (p,t)
reaction, but the angular distribution agrees with the assignment of spin $4^+$. Therefore both
levels are present in Table~\ref{tab:expEI}.

\emph{1393.4 keV}. This level was observed in the decay of $^{228}$Pa with restriction of the
spin-parity to $1^+$, $2$ and $3^-$ by its population and depopulation \cite{Web98}. Additional
restriction from the $W(90^\circ)/W(180^\circ)$ angular distribution ratio indicates that this
level has most likely $I^\pi=1^+$ and that the $2^+$ and $3^-$ assignments are nearly excluded.
This level was not observed in the (p,t) reaction, thus supporting such conclusion.

\emph{1415.8 keV.} Spin $2^+$ was assigned to this level in \cite{Dal87} and spins $2^+$ or
$3^-$ were allowed in  \cite{Web98}. The angular distribution of tritons gives preference to
spin $3^-$.

\emph{1432.1 keV.}  The discrepancy between $3^+$ in \cite{Dal87} and $4^+$ in
 \cite{Web98} for the 1432.1 keV  level is removed already by the fact of  the excitation of this state in
 the (p,t) reaction, and additionally by the angular distribution  leading to the $4^+$
 assignment. Also additional lines observed in the decay of $^{228}$Pa \cite{Web98}, leading to
 the 6$^+$ level, confirm this assignment.

\emph{1450.29 keV.} Spin $3^-$ was assigned to this level in \cite{Dal87} and spin $4^-$ in
\cite{Web98}. The fact that this level is not observed in the (p,t) reaction gives preference
for an assignment of spin $4^-$ not excluding spin $3^-$.

\emph{1531.7 keV.} Spin $3^+$ was assigned for the level at 1531.47 keV both in \cite{Dal87} and
\cite{Web98}. However, the angular distribution of tritons for the level with the close-lying
energy 1531.7 keV indicates  another spin value. It has a steeply rising cross section at small
angles as for a $0^+$ excitation, however,  the minimum at a detection angle about $14^{\circ}$
is not sharp. Therefore we assumed an overlapping of  peaks of two levels, one of which is a
$0^+$ level.

\emph{1643.8 keV.} There are two close-lying levels at 1643.18 keV with spin 2$^-$ or 3$^-$
identified both in \cite{Dal87} and \cite{Web98} and 1643.8 keV, respectively, as identified in
\cite{Web98} with an assignment of possible spins (2,3,4)$^+$. Only the level at 1643.8 keV is
observed in the (p,t) reaction with clear assignment of spin 4$^+$.

\emph{1733.8 keV.} We assumed that the level at 1735.6 keV,  identified as a 4$^+$ state in
\cite{Dal87} and as 2$^+$,3,4$^+$ state in \cite{Web98}, and the level at 1733.8 keV seen in the
(p,t) reaction with an assignment of spin 4$^+$ are identical,  though the energy difference is
larger than the energy error.

\emph{1742.8 keV.} Spin 3 was assigned to the level at 1743.86 keV in \cite{Dal87} and spin
4$^+$ in \cite{Web98}. The angular distribution from the (p,t) reaction prefers the assignment
of spin 4$^+$.

\emph{1758-1760 keV.} Several close-lying levels were identified  at 1757.9 keV with spin
$1^-,2,3^-$ \cite{Web98}, at 1758.24 keV with spin (3,4)$^+$ \cite{Dal87}, at 1760.25 keV with
spin 4$^+$ \cite{Dal87} and with spin $(2,3)^+$ \cite{Web98}. Different $\gamma$-lines were used
in the identification of the levels at 1757.9 and 1758.24 keV: 741.9, 1361.4, 1430.0 keV for the
first one and 1571.52 and 1700.59 keV for the latter. At the same time, in \cite{Dal87} the line
at 1430.0 keV was used for the identification of another level at 1617.74 keV, and the important
line at 1758.24 keV was used for the identification of the level at 1944.85 keV. The line at
1758.11 keV can be attributed to the decay of the level at 1758.24 keV, then the spin of this
state distinctly has to be $2^+$. The ambiguity cannot be solved with the (p,t) data. Therefore
we put the level at 1758.1 keV with an assignment of spin $2^+$ from the (p,t) study in
correspondence with the level at 1757.9 keV in \cite{Web98}, but for the level at 1758.24 keV in
\cite{Dal87} we do not exclude the spin $2^+$, too. As far as the level at 1760.25 keV is
concerned, a spin $2^+$ can be nearly excluded, since this line is not observed in the (p,t)
reaction.

\emph{1796.8 keV.} Two close-lying levels were identified: 1795.9 keV with an assignment
$(4^+,3^-)$ in \cite{Dal87} and  1796.4 keV with an assignment as $3^+,4,5^+$ in \cite{Web98}.
The level at 1796.8 keV with spin $4^+$ is observed in the (p,t) reaction. It is problematic to
put this level in correspondence with one of the observed ones in decay, considering the
assignments. Therefore only energetic proximity was taken into account.

\emph{1908.9 keV.} The level at 1908.4 keV, $(3^-)$ was identified in \cite{Web98}, however, a
level with almost the same energy of 1908.9 keV as observed in the (p,t) reaction was clearly
identified as a $0^+$ state,  they must be considered as a different levels.

\emph{2010.4 keV.} There is discrepancy in the assignment of spin to the level at 2010.15 keV:
$2^+,3$ in \cite{Web98} and $4^+$ in \cite{Dal87}. The angular distribution from the (p,t)
reaction prefers spin $(2^+)$.\\

\begin{figure}[h]
\begin{center}
\epsfig{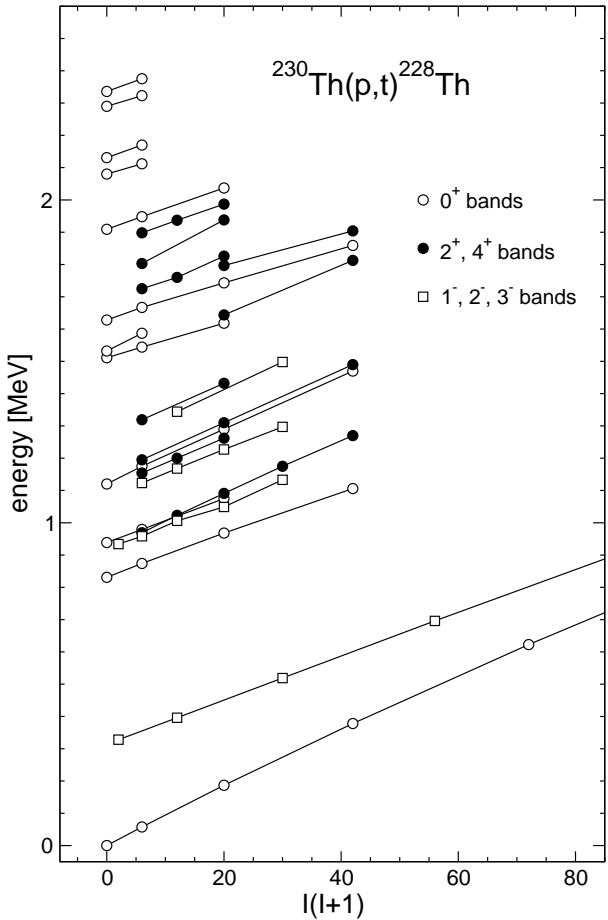} \caption{\label{fig:bands} Collective bands based on
the $0^{+}$, $2^{+}$, $1^{-}$,  $2^{-}$, and $3^{-}$ excited states in $^{228}$Th as assigned
from the DWBA fit of the angular distributions from the (p,t) reaction.}
\end{center}
\end{figure}


\begin{figure}[h]
\begin{center}
\epsfig{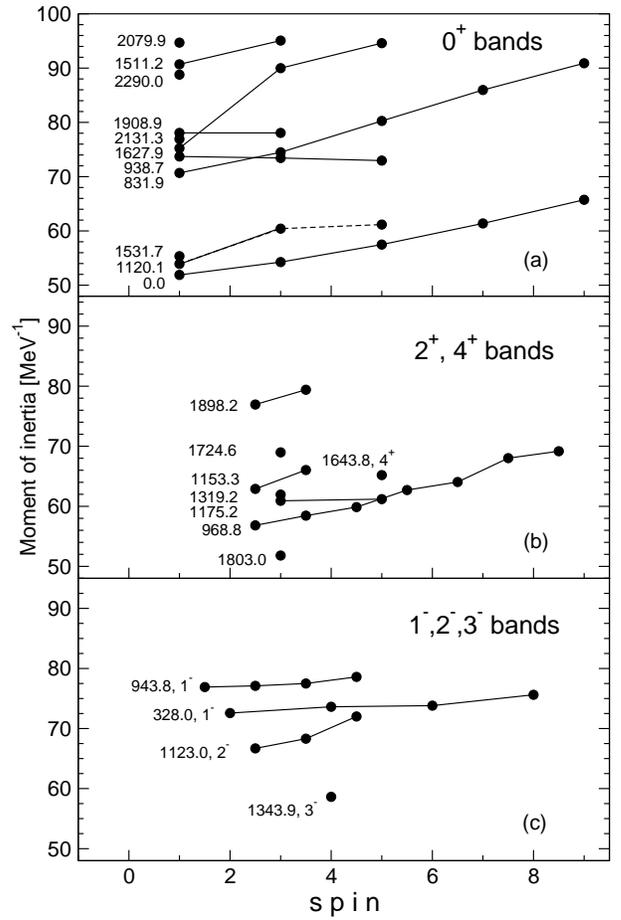} \caption{\label{fig:moments} Moments of inertia for
the bands in $^{228}$Th as assigned from the angular distributions
  from the $^{230}$Th(p,t)$^{228}$Th reaction.  Values of $J/\hbar^2$ are given.}
\end{center}
\end{figure}

\newcolumntype{d}{D{.}{.}{3}}
\begin{table*}[]
\caption{\label{tab:bands} \normalsize{The sequences of  states qualifying as candidates for
rotational bands (from the CHUCK fit, the (p,t) cross sections and the inertial parameters).
More accurate values of energies are taken from the first two columns of Table~\ref{tab:expEI}.
The energies taken in brackets correspond to the  tentatively  assigned sequences. }}
\begin{ruledtabular}
\begin{tabular}{cccccccccc}
\smallskip\\
 $0^+$ & $1^+$ & $2^+$ & $3^+$ & $4^+$ & $5^+$ & $6^+$ & $7^+$
& $ 8^+$ &
\smallskip\\
\hline
\smallskip\\
0.0 &&    57.8 &&     186.8 &&  378.2 &&   622.5 &\\
831.9 && 874.4 && 968.4 &&1105.5 &&1281  &\\
    &&   968.8 & 1022.5 & 1091.0 &1174.5 & 1270.2 &1380  &1497  & \\
    &&   1153.5 & 1200.5 & 1261.5 & &  &  &  &\\
938.6&& 979.5 && 1074.8 &&&&&\\
1120.1  && 1175.4 && 1290.2 && 1470.0& &&\\
         && 1319.2 && 1432.0 &&  &&  &\\
1511.2 && 1544.4 && 1618.3 && &&  &\\
       && 1638.3 & 1688.4 & 1760.2 &&&&&\\
       &&        && 1643.8 && (1812.7) && & \\
(1531.7) && 1586.9 &&  &&  &&&\\
1627.9 && 1667.3 && 1742.8 && 1858.6 &&&\\
     &&        &        & 1796.8 && 1903.9  &&&\\
     && 1724.6 & 1760.3 & 1826.2 &&  &&&\\
     && 1803.0 && 1938.3 &&  &&&\\
     && 1898.2 & 1937.2 & 1987.5 &&&&&\\
     && 1899.98 & 1944.85 & (2010.17) &&&&&\\
1908.9 && 1947.8 && 2037.0 &&&&&\\
2079.9 && 2111.6 &&  &&  &&&\\
2131.3 && 2170.3 &&  &&  &&&\\
2290.0 && 2323.2 &&  &&  &&&\\
(2335.9) && 2375.5 &&  &&  &&&\\

\hline
\smallskip\\
$K^\pi$& $1^-$ & $2^-$ & $3^-$ & $4^-$ & $5^-$ & $6^-$ & $7^-$ & $8^-$ &$9^-$
\smallskip\\
\hline
\smallskip\\
$0^-$& 328.0 && 396.9 && 519.2 && 695.6 && 920.6\\
$1^-$& 943.8 & 968.3 & 1016.4 & 1059.2 & 1143.2 &&&&\\
$2^-$       && 1122.9 & 1168.4 & 1226.6 & 1297.3 &&&&\\
$3^-$       &&        & 1344.0 &        & 1497.7 &&&&\\
\end{tabular}
\end{ruledtabular}
\end{table*}


\section{\label{Disc} Discussion}

\subsection{\label{sec:bands}Collective bands in $^{228}$Th}

After the assignment of spins to all excited states,  those sequences of  states can be
identified, which show the characteristics of a rotational band structure. An identification of
the states attributed to rotational bands was made on the basis of the following conditions:

a) the angular distribution for a  band member candidate state is fitted by  DWBA calculations
for the spin necessary to put this state in the band;

b) the transfer cross section in the (p,t) reaction  to states in the potential band has to
decrease with increasing  spin;

c) the energies of the states in the band can be fitted
approximately by the expression for a rotational band   $E = E_0+
AI(I+1)$ with a small and smooth variation of the inertial
parameter $A$.

 Collective bands identified in such a way are shown
in Fig.~\ref{fig:bands} and are listed in Table~\ref{tab:bands} (for a calculation of the
moments of inertia). The procedure can be justified in that some sequences meeting the above
criteria are already known from gamma-ray spectroscopy to be rotational bands
\cite{Dal87,Web98}, so similar sequences are  rotational bands, too. The straight lines in
Fig.~\ref{fig:bands} strengthen the argument for these assignments. It is worth mentioning, that
the assignments of $0^+$ even for the states at 1531.7 and 2335.9 keV are supported by one $2^+$
state on top of them. The bands built on  states of one-phonon octupole-quadruplet (the band
$K^{\pi}=1^-$ was not correctly identified in \cite{Dal87}), the band with $K^{\pi}=0^+$, 831.9
keV \cite{Dal87,Web98}, $K^{\pi}=0^+$, 938.6, 1120.1 keV \cite{Web98} and $K^{\pi}=2^+$, 968.8
and 1153.5 keV \cite{Dal87,Web98} were identified earlier. Additional levels are added to these
bands from the (p,t) study (Table~\ref{tab:bands}). Two bands with $K^{\pi}=2^+$ are added in
Table~\ref{tab:bands}, based only on the analysis of the decay of $^{228}$Ac \cite{Dal87}: at
1638.23 keV and at 1899.98 keV. There is only contradiction in the spin assignment for the
2010.17 keV level. The (p,t) data do not support spin $4^+$ and prefer a $2^+$ assignment
instead.

In Fig.~\ref{fig:moments} we present moments of inertia (MoI) obtained by fitting the level
energies of the bands displayed in Fig.~\ref{fig:bands}  by the expression $E = E_0 + AI(I+1)$
for close-lying levels,  i.e. they were determined for  band members  using the ratio of $\Delta
E$ and $\Delta [I(I+1)]$, thus saving the spin dependence of the MoI. This procedure is valid
for all bands except the 943.8 keV, $1^-$ band. The usual procedure leads to  strongly
staggering values. In the case of the $K^{\pi} = 0^-$
and $K^{\pi} = 1^-$ bands, the Coriolis interaction mixes the band members only for $I$ odd. The
$I$ even members of the $K^{\pi} = 1^-$ band remain unperturbed.
In a simple two level model ($K^{\pi} = 0^-$ and $K^{\pi} = 1^-$ bands) the following
expression can be obtained for the band energies
\begin{eqnarray}
E(I,K^{\pi}=1^-) \sim E_1 +(A_1+B)I(I+1)\enspace \makebox{for $I$ odd}  \nonumber \\
E(I,K^{\pi}=1^-) \sim E_1+A_1I(I+1) \enspace \makebox{for $I$ even}
\end{eqnarray}
where $E_1=E_1^{'}-A_1$ and $B={C^2}/({E_1^{'}-A_1-E_0^{'}})$, $E_0^{'}$ and $E_1^{'}$ are the
intrinsic bandhead energies, $A_1$ is the inertial parameter and $C$ is the strength of the
Coriolis interaction, which is believed to be small.
 An effective parameters of inertia behaves then as
\begin{eqnarray}
A_1(eff) = A_1 + \frac{1}{2}B(I+1)\enspace \makebox{for $I$ odd}  \nonumber \\
A_1(eff) = A_1 - \frac{1}{2}B(I-1)\enspace \makebox{for $I$ even}
\end{eqnarray} \\
Fitting these expressions to the experimental data gives the smoothly changing values of the
moment of inertia  between 76.9 and 78.6 as shown in Fig.\ref{fig:moments} with  parameter
$B=0.75 \div 0.68$ (thus staggering is removed).

In most cases MoI slightly increase with the increasing spin. There are few cases of $0^+$ bands
where MoI are sloping down. This can be explained as an effect of the Coriolis interaction of a
2qp $K^{\pi}=0^ +$ band with a nearby lying 2qp $K^{\pi}=1^+$ band (not seen in the (p,t)
reaction). A similar effect was observed e.g. in $^{168}$Er for the $3^-$  and $4^-$ 2qp bands
\cite{Bur85}. In $^{228}$Th an additional MoI even-odd spin staggering is expected for the
$K^{\pi}=1^+$ band similar to that for the octupole $K^{\pi}=1^-$ band since the odd spin states
of the $K^{\pi}=1^+$ band have no counterparts in the $K^{\pi}=0^+$ band.

 The obtained MoI cover a
broad range, from ~$\sim$50 MeV$^{-1}$ to ~$\sim$100 MeV$^{-1}$. The negative parity bands based
on the states with spin 1$^-$, interpreted as the octupole-vibrational bands \cite{Dal87,Web98},
have high MoI. The $0^+$ band at 1120.1 keV, considered as $\beta$ - vibrational band, has the
smallest MoI,  close to the one of the ground-state band. At this stage, it is difficult to
state a complete correlation between the intrinsic structure of the bands and the magnitude of
their MoI.  Nevertheless, one can assume for the $0^+$ bands that some of the larger MoI could
be related to the two-phonon octupole  structure and the smallest MoI could be related to the
one-phonon quadrupole structure.  The bands with intermediate values of the MoI could be based
on the two-phonon quadrupole excitations. If the moments of inertia do indeed carry information
on the inner structure of the bands, then the number of excitations with  a structure as in the
g.s. or $\beta$-vibrational states in $^{228}$Th is small.
\begin{figure}
\begin{center}
\epsfig{file=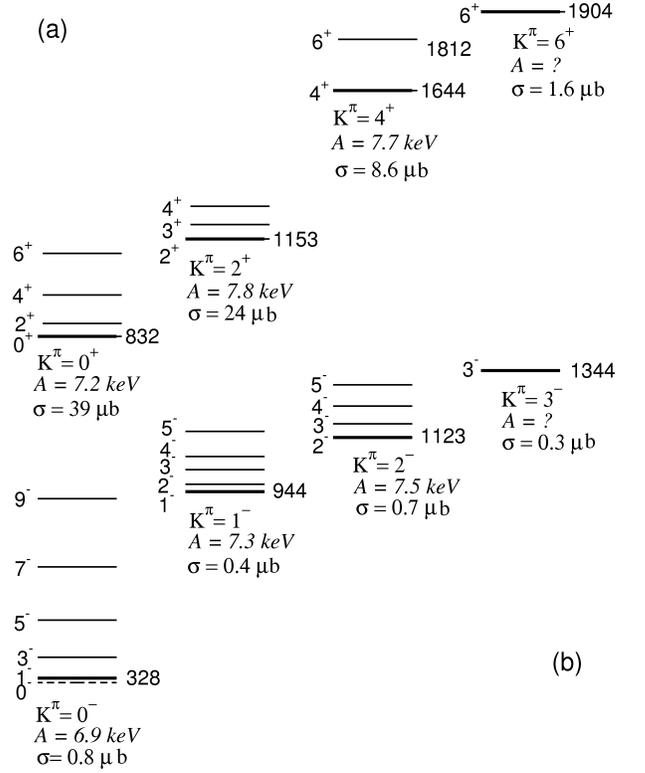, width=8 cm, angle=0} \caption{\label{fig:quadruplet} Suggested multiplets
of states of the octupole one-phonon  (bottom) and  the octupole two-phonon (top) excitations
and the corresponding collective bands.}
\end{center}
\end{figure}

\subsection{\label{sec:quadr} Quadruplets of octupole excitations}

The lowest negative-parity excitations with $K^{\pi} = 0^-, 1^-, 2^-, 3^-$ are generally
interpreted as octupole vibrational. They are one-phonon octupole excitations. The corresponding
energies in $^{228}$Th as 328.0, 944.2, 1123.0, and 1344.0 keV were established already in
\cite{Web98}. Here we confirmed these assignments after the removal of some ambiguities. They
are the bandheads of the rotational bands which are displayed in Fig.~\ref{fig:quadruplet}
together with the (p,t) cross sections and their  parameters $A$.

In the case of $^{230}$Th, the assumption was made that the strongly excited first 0$^+$ state,
together with also strongly excited states with spins 2$^+$ and 4$^+$, accompanied by somewhat weaker
excited state with spin 6$^+$,  belong to the two-phonon octupole quadruplet \cite{Lev09}.
Strong excitations and close rotational parameters  were the arguments for assigning the same
structure to these states. As one can see in  Fig.~\ref{fig:strength}, the first  0$^+$ state in
$^{228}$Th at 831.9 keV is also strongly excited. Taking into account the decay properties of
the band on this 0$^+$ state, the suggestion was made that this band has an octupole two-phonon
structure \cite{Dal87}.  The picture for other states is not so transparent. There is no
prominent excitation strength of the $2^+$ and $4^+$ states just above this 0$^+$ state. The
first excited 2$^+$ state, which is not a member of the 0$^+$ band, is the state at 968.8 keV.
But the de-excitation
 of the band built on this state demonstrates the properties expected
for a $\gamma$-vibrational
 band. Moreover, its moment of inertia is much smaller than the one
derived for the 0$^+$ band at 831.9 keV.
 For the band built on the $2^+$ state at 1153.3 keV, the moment of
 inertia is close  to the one of the band built on the $0^+$ state at
831.9 keV and the state at 1153.3
 keV is relatively strong excited in the (p,t) reaction. The $4^+$ and
$6^+$  states,
  which do not belong to rotational bands, which are strongly excited in
the (p,t) reaction and could be  members of the two-phonon octupole quadruplet, are the states
at
 1643.8 and 1905.8 keV. The level at 1812.7 keV, tentatively assigned as
a $6^+$ state, can be attributed
 to the band based on the $4^+$ state, the corresponding inertial
parameter again is very close to
 the one for  $0^+$ and $2^+$ bands. No members of a rotational band
can be related to the
  $6^+$ band head at 1905.8 keV.

\subsection{\label{sec:IBM} IBM calculations}
In the Interacting Boson Model (IBM), the positive-parity states are described by introducing
\(s\) and \(d\) bosons, while for the negative parity states one has to introduce additional
bosons with odd values of angular momentum (at least one \(f\) boson). In the region of
transitional actinides, where octupole deformation might develop, the IBM-\(spdf\) (which uses
\(p\) and \(f\) bosons) was applied with success in Refs. \cite{Eng87,Zam01,Zam03}.

In the present paper, we adopt the IBM-\(spdf\) framework for calculating the low-lying positive
and negative parity states in \(^{228}\)Th. In Ref.  \cite{Zam01}, the IBM calculations for this
nucleus have been already performed. However, these calculations used only a simplified
Hamiltonian to describe the existing (up to that date) electromagnetic data. More recent
calculations (which also used a simplified Hamiltonian) \cite{Wir04} indicated that IBM fails
completely to reproduce the (p,t) spectroscopic factors. The calculated first excited states
were found with a transfer strength of \(\simeq\)1\(\%\) of that of the ground state and the
higher states were even weaker, whereas experimentally the first excited state is seen with
\(\simeq\)30\(\%\) of the ground-state intensity. In order to treat these spectroscopic
observables in a reasonable approach, we used the method suggested in Ref. \cite{Pas10}, where
it was shown that the addition of the second-order O(5) Casimir operator in the Hamiltonian can
account for the observed (p,t) spectroscopic factors.

 The Hamiltonian employed  in the present paper is similar to the one used in Refs. \cite{Zam01, Zam03}
 and is able to describe simultaneously the positive and negative parity states:
\begin{eqnarray}
\hat{H}_{spdf}=\mathrm{\epsilon}_{d} \hat{n}_{d}+\mathrm{\epsilon}_{p}
\hat{n}_{p}+\mathrm{\epsilon}_{f} \hat{n}_{f} + \mathrm{\kappa}(\hat{Q}_{spdf}\cdot
\hat{Q}_{spdf})^{(0)}\nonumber\\
 + \mathrm{\mathit a_{3}}
[(\hat{d}^{\dagger}\tilde{d})^{(3)} \times (\hat{d}^{\dagger}\tilde{d})^{(3)}]^{(0)}\label{eq1},
\end{eqnarray}
where \(\epsilon_{d}\), \(\epsilon_{p}\), and \(\epsilon_{f}\) are the boson energies and
\(\hat{n}_{p}\), \(\hat{n}_{d}\), and \(\hat{n}_{f}\) are the boson number operators, $\kappa$
is the quadrupole-quadrupole interaction strength and $a_3$ is the strength of the O(5) second
order Casimir operator. In the \(spdf\) model, the quadrupole operator is considered as being
\cite{Kuz90}:
\begin{eqnarray}
\hat{Q}_{spdf}=\hat{Q}_{sd}+\hat{Q}_{pf}=\nonumber\\
(\hat{s}^{\dagger}\tilde{d}+\hat{d}^{\dagger}\hat{s})^{(2)}+\chi^{(2)}_{sd}(\hat{d}^{\dagger}\tilde{d})^{(2)}
+\frac{3\sqrt{7}}{5}[(p^{\dagger}\tilde{f}+f^{\dagger}\tilde{p})]^{(2)}\nonumber\\
+\chi^{(2)}_{pf} \left\{
\frac{9\sqrt{3}}{10}(p^{\dagger}\tilde{p})^{(2)}+\frac{3\sqrt{42}}{10}(f^{\dagger}\tilde{f})^{(2)}\right\}\label{eq2}
\end{eqnarray}

The quadrupole electromagnetic transition operator is:
\begin{eqnarray}
\hat{T}(E2)=e_{2} \hat{Q}_{spdf}\label{eq3},
\end{eqnarray}
where \(e_{2}\) represents the boson effective charge.
To ensure no-vanishing E2 transitions between the states containing no \(pf\) bosons and those
having \((pf)^{2}\) components we follow the approach described in Refs.
\cite{Zam01,Zam03}, where the mixing of different positive parity-states with different \(pf\)
components is achieved by introducing in the Hamiltonian a dipole-dipole interaction term of the
form:
\begin{eqnarray}
\hat{H}_{int}=\alpha \hat{D}^{\dagger}_{spdf}\cdot \hat{D}_{spdf}+ H.c.\label{eq6}
\end{eqnarray}
where
\begin{eqnarray}
\hat{D}_{spdf}=-2\sqrt{2}[{p}^{\dagger}\tilde{d}+{d}^{\dagger}\tilde{p}]^{(1)}
+\sqrt{5}[{s}^{\dagger}\tilde{p}+{p}^{\dagger}\tilde{s}]^{(1)}\\\nonumber
+\sqrt{7}[{d}^{\dagger}\tilde{f}+{f}^{\dagger}\tilde{d}]^{(1)}\label{eq7}
\end{eqnarray}
is the dipole operator arising from the \(O\)(4) dynamical symmetry limit, which does not
conserve separately the number of positive and negative parity bosons \cite{Kuz90,Kuz89}. This
term will also be important later in the calculations of the two-neutron transfer intensities.
The interaction strength is given by the \(\alpha\) parameter and is chosen to have a very small
value, \(\alpha\)=0.0005 MeV, similar to Refs. \cite{Zam01,Zam03}.

For the \(E1\) transitions, a linear combination of the three allowed one-body interactions was
taken:
\begin{eqnarray}
\hat{T}(E1)=e_{1}[\chi_{sp}^{(1)}({s}^{\dagger}\tilde{p}+{p}^{\dagger}\tilde{s})^{(1)}
+({p}^{\dagger}\tilde{d}+{d}^{\dagger}\tilde{p})^{(1)}\nonumber\\
+\chi_{df}^{(1)}({d}^{\dagger}\tilde{f}+{f}^{\dagger}\tilde{d})^{(1)}]\label{eq4},
\end{eqnarray}
where \(e_{1}\) is the effective charge for the \(E1\) transitions and \(\chi_{sp}^{(1)}\) and
\(\chi_{df}^{(1)}\) are  model parameters.

\begin{figure*}
\centering
\includegraphics[width=14cm, angle=0]{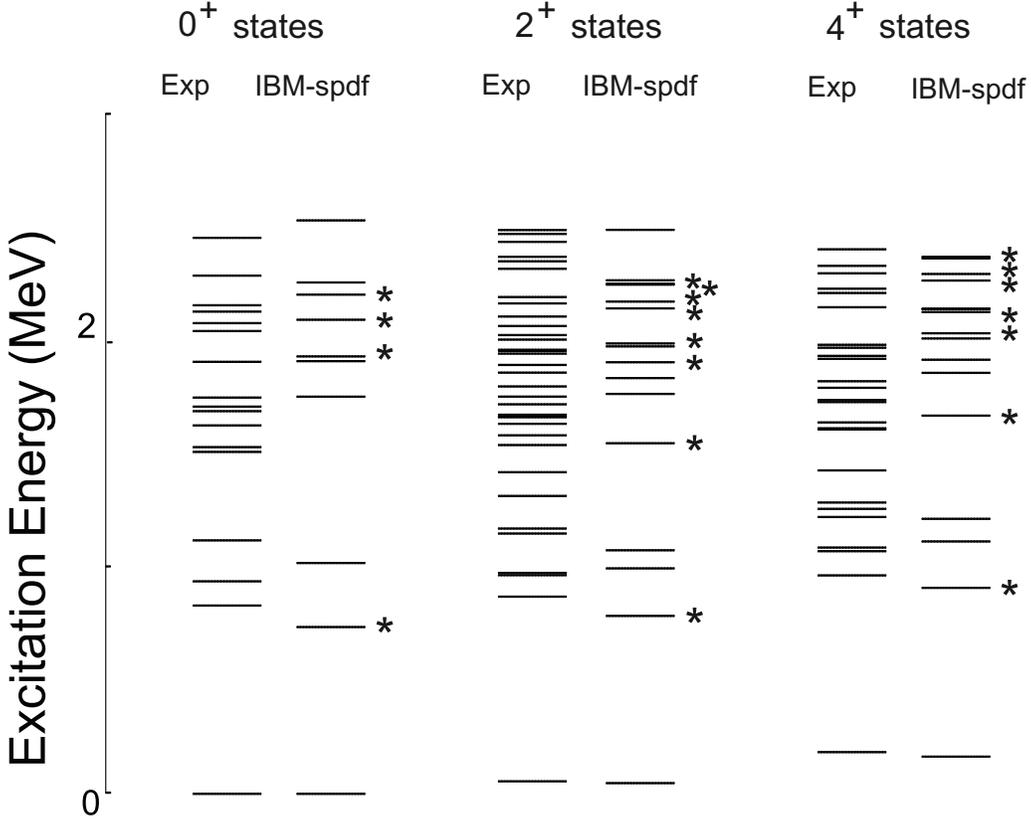}
\caption{\label{fig10} Energies of all experimentally assigned excited 0\(^{+}\), 2\(^{+}\), and
4\(^{+}\) states in \(^{228}\)Th in comparison to IBM-\(spdf\) calculations. For the 0\(^{+}\)
excitations, the states containing 2 \(pf\) bosons in their structure and assumed to have a
double dipole/octupole character are marked with asterisk.}
\end{figure*}

The goal of the present paper is to describe simultaneously both the existing electromagnetic
and the hadronic (transfer strength) data. To achieve this goal, two-neutron transfer
intensities between the ground state of the target nucleus and the excited states of the
residual nucleus were also calculated. The L=0 transfer operator may contain various terms, but
we shall restrict our operator to the following form:

\begin{eqnarray}
\hat{P}^{(0)}_{\nu}=(\alpha_{p}\hat{n}_{p}+\alpha_{f}\hat{n}_{f})\hat{s}+\nonumber\\
+\alpha_{\nu} \left (\Omega_{\nu}-N_{\nu}-\frac{N_{\nu}}{N} \hat{n}_{d}\right)^{\frac{1}{2}}
\left(\frac{N_{\nu}+1}{N+1}\right)^{\frac{1}{2}} \hat{s}\label{eq5},
\end{eqnarray}
where \(\Omega_{\nu}\) is the pair degeneracy of neutron shell, \(N_{\nu}\) is the number of
neutron pairs, \(N\) is the total number of bosons, and \(\alpha_{p}\), \(\alpha_{f}\), and
\(\alpha_{\nu}\) are constant parameters. The L=0 transfer operator contains two additional
terms beside the  leading order term, proportional to the bosonic \(\hat{s}\) operator
\cite{Iach87}. Details about the contributions of different terms in calculating the (p,t)
spectroscopic factors will be given in a forthcoming paper \cite{Pas_unp}.

The calculations were performed using the computer code OCTUPOLE \cite{Kuz_comp}. The
Hamiltonian is diagonalized in a Hilbert space with a total number of bosons
\(N_{B}=n_{s}+n_{d}+n_{p}+n_{f}\). For the present calculations we used an extended basis
allowing up to three negative parity bosons (\(n_{p}+n_{f}\)=3). The vibrational strengths used
in the calculations are \(\epsilon_{d}\)=0.2 MeV, \(\epsilon_{p}\)=1.0 MeV, and
\(\epsilon_{f}\)=1.1 MeV, while the quadrupole-quadrupole interaction strength has a value of
\(\kappa\)=-21 keV. The strength of the O(5) second order Casimir operator is set to
\(a_{3}\)=0.053 MeV, while the quadrupole operator parameters are (\(\chi^{(2)}_{sd}\)=-1.09,
\(\chi^{(2)}_{pf}\)=-1).

The full spectrum of excited 0\(^{+}\) states obtained in the present experiment is displayed in
Fig.~\ref{fig10} in comparison to the corresponding calculated values.     In the energy range
covered experimentally (up to 2.5 MeV), the IBM-\(spdf\) calculations predict 10 excited
0\(^{+}\) states in comparison to the 17 experimentally observed 0\(^{+}\) excitations (firm
spin assignment). Given that there was no attempt to fine tune the calculations to the empirical
0\(^{+}\) states, there is no point in invoking a precise energy cutoff for the IBM
calculations. Therefore, it is appropriate to look also above 2.5 MeV, where there is a
continuous spectrum of  0\(^{+}\) states consisting of 20 states up to 3.3 MeV and as many as 30
up to 4 MeV. The IBM predicts that some of these states are having 2\(pf\) bosons in their
structure and are related (according to Ref. \cite{Zam01}) to the presence of double
dipole/octupole excitations.  For example, the boson admixtures for the first excited $0^+$
state are $n_d$=4.2, $n_p$=1.4, $n_f$=0.6 in comparison with those for the $\beta$-vibrational
state as $n_d$=4.5, $n_p$=0.006, $n_f$=0.001. However, the present data cannot allow for a final
decision on the nature of the 0\(^{+}\) states. Additional experimental information is needed to
measure the branching ratios and the absolute transition probabilities stemming from these
states. In Fig.~\ref{fig10}, the 2\(^{+}\) and 4\(^{+}\) levels revealed in the present
experiment are also compared to the predictions of the IBM. The experiment revealed 33 excited
2\(^{+}\) and 25 excited 4\(^{+}\) states up to 2.5 MeV. In the same energy range, the
calculations produced only 16 excited 2\(^{+}\) states and 15 4\(^{+}\) excitations. If one
looks above this limit, the IBM predicts 32 excited 2\(^{+}\) states and 33 excited 4\(^{+}\)
states up to 3.3 MeV.

\begin{table}
\caption{\label{BE_IBM} Experimental and calculated B(E1)/B(E2) transition ratios in
\(^{228}\)Th. The parameters of the \(E1\) operator are fitted to the experimental data
available.}
\begin{ruledtabular}
\begin{tabular}{ccccccc}
K$^{\pi}$ & E$_{i}$ (keV) & J$_{i}$ & J$_{f1}$ & J$_{f2}$ & Exp. (10$^{-4}$ b$^{-1}$) & IBM (10$^{-4}$ b$^{-1}$)\\
\hline
0$^{+}_{1}$ & 832  & 0$^{+}$ & 1$^{-}_{1}$ & 2$^{+}_{1}$  & 5.1(4)       & 6.1     \\
                       & 874  & 2$^{+}$ & 3$^{-}_{1}$ & 4$^{+}_{1}$  & 7.1(15)     & 7.6      \\
                       &         & 2$^{+}$ & 3$^{-}_{1}$ & 2$^{+}_{1}$  & 24.5(31)   & 15.2   \\
                       &         & 2$^{+}$ & 3$^{-}_{1}$ & 0$^{+}_{1}$  & 14.7(24)   & 23.6   \\
                       &         & 2$^{+}$ & 1$^{-}_{1}$ & 4$^{+}_{1}$  & 4.2(9)       & 4.4     \\
                       &         & 2$^{+}$ & 1$^{-}_{1}$ & 2$^{+}_{1}$  & 14.5(19)   & 8.9     \\
                       &         & 2$^{+}$ & 1$^{-}_{1}$ & 0$^{+}_{1}$  & 8.7(14)     & 13.7   \\
                       & 968  & 4$^{+}$ & 5$^{-}_{1}$ & 6$^{+}_{1}$  & 22.8(80)   & 9.2     \\
                       &         & 4$^{+}$ & 5$^{-}_{1}$ & 4$^{+}_{1}$  & 10.8(27)   & 18.2   \\
                       &         & 4$^{+}$ & 5$^{-}_{1}$ & 2$^{+}_{1}$  & 6.7(13)     & 20.7   \\
                       &         & 4$^{+}$ & 3$^{-}_{1}$ & 6$^{+}_{1}$  & 19.1(67)   & 5.9     \\
                       &         & 4$^{+}$ & 3$^{-}_{1}$ & 4$^{+}_{1}$  & 9.0(23)     & 11.8   \\
                       &         & 4$^{+}$ & 3$^{-}_{1}$ & 2$^{+}_{1}$  & 5.6(11)     & 13.4   \\
0$^{+}_{3}$ & 1176& 2$^{+}$ & 1$^{-}_{1}$ & 4$^{+}_{1}$  & 0.060(25) & 0.08   \\
                       &         & 2$^{+}$ & 1$^{-}_{1}$ & 2$^{+}_{1}$  & 0.25(10)   & 0.27   \\
                       &         & 2$^{+}$ & 1$^{-}_{1}$ & 0$^{+}_{1}$  & 0.62(28)   & 0.38   \\
                       &         & 2$^{+}$ & 3$^{-}_{1}$ & 4$^{+}_{1}$  & 0.09(5)     & 0.16   \\
                       &         & 2$^{+}$ & 3$^{-}_{1}$ & 2$^{+}_{1}$  & 0.39(20)   & 0.52   \\
                       &         & 2$^{+}$ & 3$^{-}_{1}$ & 0$^{+}_{1}$  & 0.95(51)   & 0.72   \\
\end{tabular}
\end{ruledtabular}
\end{table}

 In \(^{228}\)Th there are no lifetimes measured for the negative-parity states, hence no absolute
transition probabilities could be extracted. Therefore we would restrict the present discussion
to reproducing the B(E1)/B(E2) ratios. A detailed comparison between the experimental data and
the present calculations is presented in Table \ref{BE_IBM}. The agreement is obtained by using
\(e_{1}\)=0.005 \(e\)fm and \(e_{2}\)=0.19 \(e\)b as the effective charges in Eq.(\ref{eq4}) and
(\ref{eq3}), respectively. The remaining \(E1\) parameters are \(\chi_{sp}\)=0.4 and
\(\chi_{df}\)=-1.4.

The B(E1)/B(E2) ratios discussed in Table \ref{BE_IBM} belong to the $K^{\pi}$=0\(_{1}^{+}\)
(the predicted double-octupole phonon band) and $K^{\pi}$=0\(_{3}^{+}\) (\(\beta\)-vibrational)
bands. The comparison between them is important, because it can be used as a tool for providing
additional information about the nature of the $K^{\pi}$=0\(_{1}^{+}\) band. All the states
belonging to this band are having 2 \(pf\) bosons in their structure in the IBM calculations and
are supposed to have a double-octupole phonon character. Further information confirming this
hypothesis comes from the analysis of the \(E1\) and \(E2\) branching ratios. If the picture
proposed by the IBM is correct, the states belonging to the $K^{\pi}$=0\(_{1}^{+}\) band will
show strong transitions into the negative-parity states
 (if they have a double-octupole character), while the levels stemming from the $K^{\pi}$=0\(_{3}^{+}\) (\(\beta\)-band)
  will show very weak \(E1\) transitions to these states. The experimental values in Table \ref{BE_IBM}
  fully confirm this hypothesis, showing that the B(E1)/B(E2) ratios are at least one order of magnitude
  larger for the $K^{\pi}$=0\(_{1}^{+}\) band.

\begin{figure}
\centering
\includegraphics[width=8 cm, angle=0]{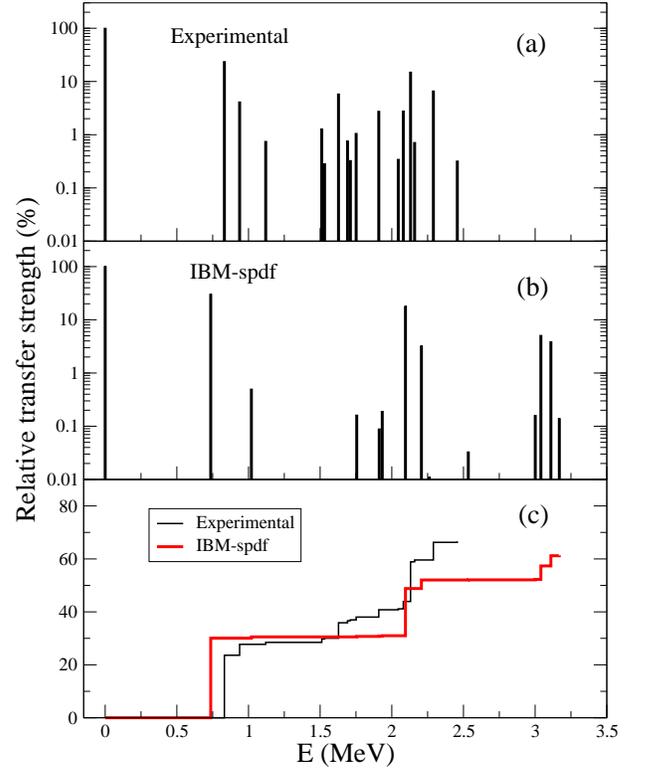}
\caption{\label{fig11} (Color online) Comparison between the experimental  two-neutron transfer
intensities (panel (a)) for the 0\(^{+}\) states and the IBM predictions (panel (b)). In panel
(c) the experimental versus computed running sum of the (p,t) strengths is given.}
\end{figure}

In Fig. \ref{fig11}, we display the calculated two-neutron intensities for \(^{228}\)Th in
comparison to the integrated experimental cross sections normalized to that of the ground state.
 The calculations reproduce the strong excitation of the first 0$^+$ state at 832 keV in
good agreement with the experiment. The experimental spectrum of 0\(^{+}\) states is dominated
also  by a single state located at an energy of 2.1 MeV, showing a high cross section of about
15\(\%\) of that of the ground state. In the IBM, there is predicted a state located at 2.1 MeV,
which have the transfer intensities of about 18\(\%\). This state has a double-octupole phonon
structure. Another state at 2.29 MeV with a relative cross section of about 7\(\%\) can be put
in correspondence to the predicted in the IBM state at 2.2 MeV also with a double-octupole
phonon structure. The running sum in Fig.~\ref{fig11} is taken up to 3.25 MeV, where another
group of states with significant transfer strength is predicted by the IBM. The parameters from
Eq.(\ref{eq5}) were estimated from the fit of the known two-neutron transfer intensities
(integrated cross sections) in Table~\ref{tab:expEI}.
 The values employed in the present paper are \(\alpha_{p}\)=1.3 mb/sr,
 \(\alpha_{f}\)=-0.4 mb/sr, and \(\alpha_{\nu}\)=0.03 mb/sr.
The location and transfer intensity of the strongest states
  is very well reproduced by the calculations. Because the calculated energy distribution
 of the 0\(^{+}\) states is underestimating the experimental data, this also affects
the fragmentation of the transferred strength. However, the main characteristics
 are well reproduced by the present calculations.

\subsection{\label{sec:QPM} QPM calculations}

 The IBM is a phenomenological approach. To gain a detailed information on the properties of
 the states excited in the (p,t) reaction, a microscopic approach is necessary.
 The ability of the QPM to describe multiple
$0^+$ states (energies, $E2$ and $E0$ strengths, two-nucleon spectroscopic factors) was
demonstrated for $^{158}$Gd \cite{LoI04}. An extension of the QPM to describe the $0^+$ states
in the actinides \cite{LoI05} was made  after our publication on the results of  a preliminary
analysis of the experimental data \cite{Wir04}. These calculations are used to compare to the
present detailed analysis of the experimental data for $^{228}$Th. As for the theoretical basis
of the calculations, we refer to the publications \cite{LoI05,Sol92}.

\begin{figure}
\centering \epsfig{file=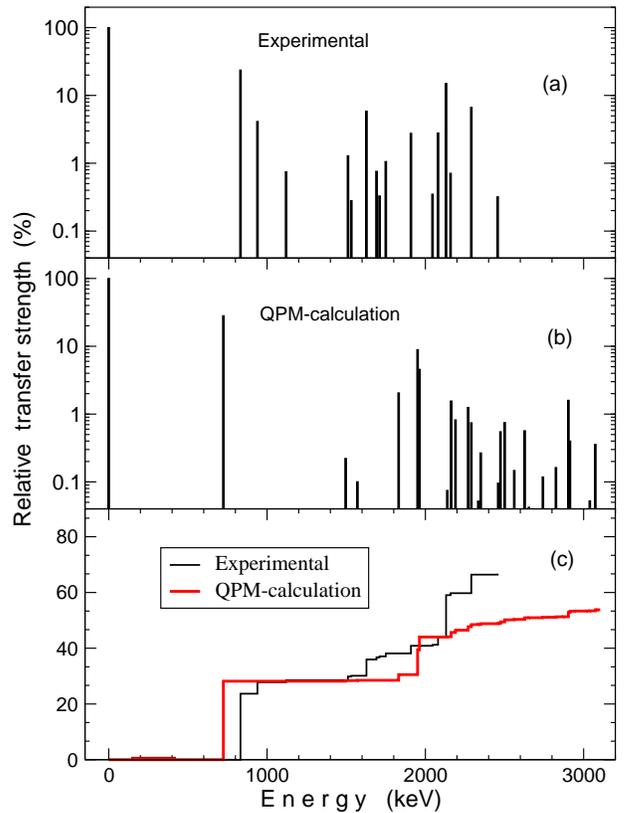, width=8 cm, angle=0} \caption{ \label{fig:compil-zeroQPM}
(Color online) Comparison of experimental and calculated
    (QPM) 0$^+$ relative level  reaction  strengths for the (p,t)
    reaction (two upper panels). The values for the 0$^+_{g.s.}$ are normalized to 1.
    The experimental increments of the (p,t) strength (absolute values) in comparison to
    the QPM calculations are shown in lower panel.}
\end{figure}

The experimental spectrum of the $0^{+}$ relative level reaction strength for the (p,t) transfer
(the ratios of the (p,t) strength for every state to those for the ground state)
 is compared to the results of the QPM calculations in Fig.~\ref{fig:compil-zeroQPM}.
The (p,t) normalized transfer
spectroscopic strengths in the QPM are expressed also as ratios
\begin{equation}\label{spec-fact}
S_n(p,t) = \left[\frac{\Gamma_n(p,t)}{\Gamma_0(p,t)}\right]^2.
\end{equation}
The amplitude $\Gamma_0(p,t)$ refers to the transitions to the $I$ members of the ground-state
rotational band, i.e. to the ground state at an analysis of the $0^+$ excitations. The amplitude
$\Gamma_n(p,t)$ includes the transitions between the ground state and the one- and two-phonon
components of the wave function.
The numerical results of the calculations obtained according to the QPM investigation in the
publication \cite{LoI05} are provided to us by A.~V.~Sushkov \cite{Sush}.
  The QPM generates 15 $0^+$ states below 2.5 MeV,  in fair agreement with the 17 firmly assigned states.
  The calculations
 reproduce the strong excitation of the first 0$^+$ state in accordance with the experiment.
 In Fig.~\ref{fig:compil-zeroQPM}, we present also the increments
 of the (p,t) strength ratios in comparison to those of the QPM calculated
 normalized spectroscopic strengths.
 As one can see, the calculations are in fair agreement with the experiment.
 The (p,t) strength for the questionable 0$^+$ state at 2335.9 keV does not influence
 considerably the results of comparison (hence it is not included in the comparison).

 Visible deviation of calculated strength from experiment is seen
 above 2 MeV. Theory predicts many $0^+$ excitations at higher energies (more than 80 states
 in the energy range below 4 MeV), but with small strength. At the same time, two
 strong excitations are observed in the experiment at 2.13 and 2.29 MeV (see Table~\ref{QPM_structure}),  respectively.
 It is interesting to note that both the IBM and the QPM  predict two strongly excited states and therefore a
 jump in the increments of the (p,t) strength
 in the vicinity of 2 MeV, thus reproducing partly the sharp increase in the experimental increment.

 In these QPM calculations, the dominant phonon structure of the 0$^+$ states in low part of energies  is
 the one-phonon quadrupole nature. For higher energies admixtures
 of two quadrupole and two octupole phonons are present in the structure of these states, and for some of the states
they became  dominant. The relatively modest role of the octupole  phonons
 in the structure of the low energy  $0^+$ states is explained in \cite{LoI05} by the enforcement of the Pauli
 principle, leading to a spreading of the lowest  two octupole  phonon components among several QPM
 $0^+$ states and  pushing them to higher energies.

\begin{table*}
 \caption{\label{QPM_structure} Phonon structure of selected QPM states.
The weights of the one-phonon ($|\lambda\mu|_i$) or the two-phonon
$(|(\lambda\mu)_i(\lambda\mu)_i|)$ components are given in percent.  Only  main one-phonon
components are shown. Transfer factors $S(p,t)$ are normalized to the 0$^+_{g.s.}$ state.}
\begin{ruledtabular}
\begin{tabular}{ccccccl}
$K^{\pi}_n$ & $E_{n}(exp)$  & $E_{n}(calc)^*$ &  $E_{n}(calc)^{**}$ & $S(p,t)_{exp}$ & $S(p,t)_{calc}^{**}$ & ~~Structure from \cite{LoI05} \hspace{35mm} Structure from\cite{Web98} \\
\hline
\vspace{0.02mm}\\
0$_1^{+}$   &   0.832       & 0.8            &   0.724          &  0.236         & 0.281 & $(20)_196$ \hspace{50mm} $(20)_197;[(30)_1(30)_2]0.3$       \\
0$_2^{+}$   &   0.939       & 1.0            &   1.496          &  0.041         & 0.002 & $(20)_294;(20)_34$ \hspace{28mm}$(20)_295;(20)_10.8;[(30)_1(30)_1]0.7$       \\
0$_3^{+}$   &   1.120       & 1.2            &   1.570          &  0.008         & 0.001 & $ (20)_24;(20)_393$ \hspace{39.8mm} $(20)_382;(20)_414;(20)_21$      \\
0$_4^{+}$   &   1.511       & 1.4            &   1.831          &  0.013         & 0.021 & $(20)_455;(20)_58;(20)_612;[(30)_1(30)_1]21$ \hspace{18mm}$(20)_478;(20)_317$      \\
0$_5^{+}$   &   1.532       &               &                   &  0.003         &       & \\
0$_6^{+}$   &   1.628       & 1.6            &   1.950          &  0.058         & 0.089 & $(20)_420;(20)_573;(20)_63$\hspace{20mm}$(20)_597;(20)_41;[(30)_1(30)_1]0.5$     \\
0$_7^{+}$   &   1.691       & 1.7            &   1.962          &  0.008         & 0.046 & $(20)_415;(20)_512;(20)_662;[(30)_1(30)_1]6$\hspace{12mm}$(20)_692;[(30)_1(30)_1]4$      \\
0$_8^{+}$   &   1.710       & 1.8            &   2.138          &  0.003         & 0.001 & $(20)_789;[(22)_1(22)_1]4$\hspace{33.5mm}$(20)_797;[(30)_1(30)_2]0.2$      \\
0$_9^{+}$   &    1.750      &               &                   &  0.011         &       & \\
0$_{10}^{+}$   &   1.909       & 1.9            &   2.162          &  0.028      & 0.016 & $(20)_69;(20)_853;[(22)_1(22)_1]4;[(30)_1(30)_1]23$\hspace{3.5mm}$(20)_874;(20)_94;(20)_64$      \\
0$_{11}^{+}$   &   2.045       &                &   2.190          &  0.003      & 0.008 & $(20)_76;[(22)_1(22)_1]87$      \\
0$_{12}^{+}$   &   2.080       &                &   2.270          &  0.028      & 0.013 & $(20)_43;(20)_65;(20)_837;(20)_914;(20)_{10}3;[(30)_1(30)_2]32$  \\
0$_{13}^{+}$   &   2.131       &                &   2.290          &  0.150      & 0.008 & $(20)_85;(20)_980;[(30)_1(30)_1]6;[(31)_1(31)_1]7$\\
0$_{14}^{+}$   &   2.159       &                &   2.334         &  0.007       & 0.001 & $(20)_94;[(30)_1(30)_2]87$      \\
0$_{15}^{+}$   &   2.290       &                &   2.350         &  0.067       & 0.003 & $(20)_{10}2;(20)_{14}30;[(22)_1(22)_2]65$      \\
0$_{17}^{+}$   &   2.456       &                &   2.359          &  0.003      & 0.001 & $(20)_{10}84;[(30)_1(30)_1]3;[(30)_1(30)_3]3$\\
\hline
0$_1^{-}$   &   0.328       & 0.5            &                     &  0.005         &       & \hspace{65mm}$(30)_199 $      \\
1$_1^{-}$   &   0.944       & 1.0            &                     &  0.002         &       & \hspace{65mm}$(31)_198 $      \\
3$_1^{-}$   &   1.344       & 1.4            &                     &  0.002         &       & \hspace{65mm}$(33)_195;~[(20)_1(33)_1]3 $      \\
\hline
2$_1^{+}$   &   0.969       & 1.0            &                     &  0.121         &       & \hspace{65mm}$(22)_198 $      \\
2$_2^{+}$   &   1.153       & 1.3            &                     &  0.145         &       & \hspace{65mm}$(22)_299 $      \\
4$_1^{+}$   &   1.432       & 1.5            &                     &  0.001         &       & \hspace{65mm}$(44)_199 $      \\
\vspace{0.02mm}
\end{tabular}
\end{ruledtabular}
$^*$ Data are taken from \cite{Web98}.  \hspace{10mm} $^{**}$ Data are taken from \cite{LoI05}\\
\end{table*}

\begin{table*}
\caption{\label{BE_QPM} Experimental and calculated B(E1)/B(E2) transition ratios in
\(^{228}\)Th. The Weisskopf estimate of this ratio is B(E1)/B(E2)= 2.9 b$^{-1}$ to be compared
to listed values.}
\begin{ruledtabular}
\begin{tabular}{crccrccrcccc}
K$^{\pi}_i$ & E$_{i}$  & I$_{i}$ && E$_{f1}$ & I$_{f1}$ && E$_{f2}$ & I$_{f2}$ && Exp. (10$^{-4}$ b$^{-1}$) & QPM (10$^{-4}$ b$^{-1}$)\\
\hline \vspace{0.02mm}\\
0$^{+}$ & 831.8        & 0$^{+}$ &&328.0 & 1$^{-}$ &&57.8 & 2$^{+}$  && 5.1(4)       & 2.25     \\
0$^{+}$ & 938.6        & 0$^{+}$ &&328.0 & 1$^{-}$ &&57.8 & 2$^{+}$  && 6.7(6)       & 54.8     \\
0$^{+}$ & 1175.5       & 2$^{+}$ &&328.0 & 1$^{-}$ &&186.8 & 4$^{+}$  && 0.06(3)       & 1.45     \\
\vspace{0.02mm}
\end{tabular}
\end{ruledtabular}
\end{table*}

  Besides the publication \cite{LoI05}, other calculations of the same Dubna group
 were  carried out for a microscopic description of the level structure, and transition
  rates between excited states in \(^{228}\)Th, observed in the decay of \(^{228}\)Pa \cite{Web98}.
  The wave functions,
 the level energies  from
 two publication  in correspondence to the experimental ones and to the transfer factors
 are given in Table~\ref{QPM_structure}. The transfer factors and
  also the moments of inertia are taken into account to put in correspondence the experimental
  and calculated levels. The large difference in the transfer factors is seen only for two levels
  at 2131 and 2290 keV.
 There is an  essential
 difference in the energies of the lowest $0^+$ states in two publications,
 as they are more close to the experimental
 ones in \cite{Web98}.  This is caused by the choice of the
isoscalar quadrupole-quadrupole interaction strength stronger than the critical value in Ref.
\cite{Web98}. As a consequence the energy of the lowest collective $0^+$ state becomes imaginary
and its properties such as the structure, E2 reduced transition probabilities and the transfer
factor are partially transferred to the next $0^+$ collective state. Among the transition
properties, the calculated B(E1)/B(E2) ratios at the decay
 of some 0$^+$ states are of special interest  in this publication. We present here some data on
 the B(E1)/B(E2) ratios,  in order to note the difference in the explanation of the experimental
 data by the QPM and the IBM.
 The calculated ratios for
transitions from the $0^+_1$,  $0^+_2$, and $0^+_3$ states to the octupole  $0^-_1$ state and
the ground state band are compared to experimental ratios in Table~\ref{BE_QPM}. As one can see
from Table~\ref{QPM_structure} and Table~\ref{BE_QPM}, the small admixtures of the octupole
two-phonon components to the wave functions of the $0^+_1$ and $0^+_2$ states are responsible
for the  fast $E1$ transitions. In the IBM, similar results are obtained for the $0^+_1$ state,
having 2 \(pf\) bosons in their structure and which are supposed to have mainly a
double-octupole phonon character (see Table~\ref{BE_IBM} and corresponding discussion). At the
same time, the B(E1)/B(E2) ratio is considerably smaller for the decay of the $0^+_3$ state,
which is a $\beta$-vibrational state, again in agreement with the experiment. The same result is
obtained in the IBM.

Generally, the QPM is quite accurate in nuclei with small ground-state correlations. These
correlations increase with the collectivity of the first one-phonon states, which is exactly the
case of the $K=0^-$ octupole phonon state in $^{228}$Th.  To decrease the correlations
  the value of the octupole-octupole isoscalar interaction strength was diminished
so that the calculated energy of the $K=0^-$ state was almost 200 keV higher than the
experimental value \cite{Web98}. In addition, the effect of multi-phonon admixtures (three and
more phonons) that pushes two-octupole phonon poles and consequently two-octupole phonon
energies to lower values was then underestimated. In summary, the accuracy of the calculations
of $^{228}$Th as stated in Ref. \cite{Web98} is worse due to the increased ground-state
correlations and the shift of two-phonon poles towards smaller energies. In future QPM studies
one also has to take into account the spin-quadrupole interaction that is known to increase the
density of low-lying $0^+$ states \cite{Pya67, Sol76}.

\subsection{\label{sec:0_concl} To the nature of $0^+$ excitations}

At a microscopic approach there can be a few situations of structure of the $0^{+}$ states. A
$\beta$-vibrational mode can be characterized by the relatively small two-nucleon transfer
strength and a relatively large B(E2) value with a moment of inertia close to the one of the
ground state. The large ratio B(E1)/B(E2) and the increase of the moment of inertia indicate the
presence of the octupole two-phonon component. If a state has a relatively weak B(E2) value and
also a weak two-nucleon transfer strength, but exhibits an increase of the moment of inertia, it
should be a state with one dominant 2qp configuration. The pairing vibrational excitations can
be characterized by their large two-nucleon transfer strengths and relatively small B(E2)
values.

It is clear that the first and second $0^+$ excited states cannot be the $\beta$-vibrational
states as usually is observed in deformed rare-earth nuclei since their moments of inertia are
much larger than those of the ground state and also their (p,t) strengths are large. The actual
$\beta$-vibrational state is observed at 1120 keV and it is excited very weakly in the (p,t)
reaction. As we have seen, both the IBM and the QPM reproduce the $0^{+}$ relative level
reaction strength for the (p,t) reaction and the B(E1)/B(E2) ratios of the decay of the lowest
$0^+$ states reasonably well. At the same time, the nature of $0^+$ excitations  in the QPM
differs significantly from the one in the {\it spdf}-IBM. In all low-lying states of the QPM
calculations, quadrupole phonons are dominant and the octupole phonons are predicted to play a
relatively modest role, whereas the IBM predicts the lowest $0^+$ state as having mainly 2\(pf\)
bosons in their structure \cite{Zam01}. The analysis of the lowest quadrupole phonon wave
function in the QPM reveals that the backward RPA amplitudes $\varphi$ contribute considerably
to the relative (p,t) reaction strength thus indicating that the lowest excited $0^+$ state
describe pairing vibrations arising from ground state fluctuations \cite{LoI05}.

\begin{figure}
\begin{center}
\epsfig{file=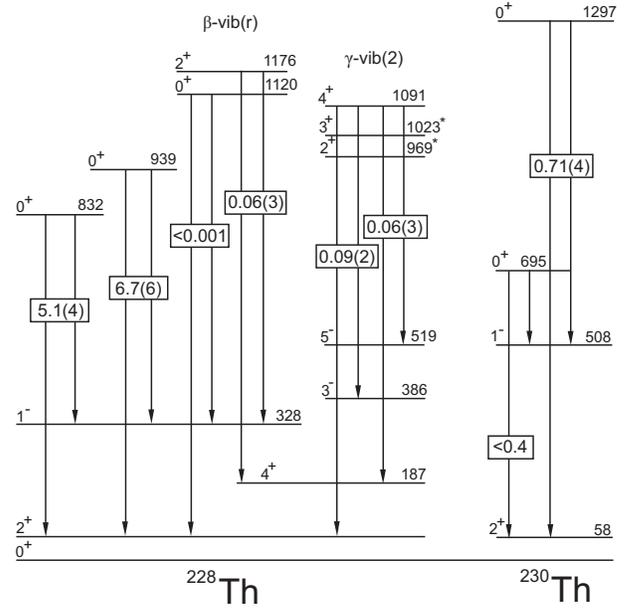, width=8 cm, angle=0} \caption{\label{fig:B(E1)-B(E2)} Comparison of the
B(E1)/B(E2) values for the lowest excited $0^+$ levels in $^{228}$Th and $^{230}$Th}.
\end{center}
\end{figure}

For an additional hint to the nature of the lowest $0^+$ states, we include
Fig.~\ref{fig:B(E1)-B(E2)} with the B(E1)/B(E2) ratio stemming from the lowest excited states in
$^{228}$Th and $^{230}$Th. Intuitively, one would expect that a large B(E1)/B(E2) ratio might be
characteristic for a two-octupole-phonon excitation, whereas a small ratio might indicate a
shape oscillation. Such a picture is observed for $^{228}$Th: large B(E1)/B(E2) ratios for the
$0^+_1$ and $0^+_2$ states, and vanishing values for $\beta$- and $\gamma$-vibrational states.
The assumption in \cite{Gra03} that just the two-phonon structure is a reason of the very strong
excitation of the $0^+_1$ state in the (p,t) reaction has to be rejected, since the B(E1)/B(E2)
ratio for the $0^+_1$ state in $^{230}$Th is small at strong excitation in the (p,t) reaction
(or  the B(E1)/B(E2) ratio is no reliable manifestation of the two-phonon nature of a state).

In this aspect a comparison of the relative transfer strengths in the (p,t) reaction leading to
the $0^+$ states in $^{228}$Th and the $3/2^-$ states in $^{229}$Pa (data are taken from
\cite{Lev94}) has to be considered additionally (see Fig.~\ref{fig:B(E1)-B(E2)}). Since the
ground-state spin in the target nucleus $^{231}$Pa  is $3/2^-$, just these spins are excited in
a two-neutron $L=0$ transfer. $^{229}$Pa can be regarded as $^{228}$Th plus a strongly coupled
proton. The rotational band built on the first $3/2^-$ excited state in $^{229}$Pa at 11.6 keV
\cite{Lev98} corresponds to the g.s. band in $^{228}$Th and is excited very strongly. The main
component in the structure of the ground state in the target nucleus $^{231}$Pa is 1/2[530]
\cite{Lev94}. Therefore the levels excited in the $L=0$ transfer are $I^{\pi},K = 3/2^-,1/2$
states and they are members of collective bands based on the state originating from coupling the
$K^{\pi} = 1/2^-$ state to the $^{228}$Th core-excited states. Such bands, as identified in
\cite{Lev94}, can be used to derive the moments of inertia for at least three $3/2^-$ states.
Values of $J/\hbar^2$ in MeV$^{-1}$ are
given below  (energies in $^{229}$Pa are relative to the lowest $3/2^-$ state) \\
\begin{tabbing}
$^{229}$Pa:  {\bf 79.2}(0.0 keV), {\bf 78.4}(703 keV), {\bf 127.5}(830 keV)\\
           \hspace{10mm} {\bf 89.3}(1524 keV). \\
$^{228}$Th: {\bf 51.9}(0.0 keV), \= {\bf 70.3}(832 keV), \= {\bf 73.2}(939 keV).\\
\end{tabbing}
The moments of inertia in $^{229}$Pa are larger than those in $^{228}$Th which is a
manifestation of the contribution of the odd proton. The large moment of inertia for the 830.5
keV state can probably be explained (at least partly) by neglecting  the Coriolis coupling when
fitting the energies of the corresponding band. Nevertheless the increment of the moment of
inertia for the state at 830.5 keV relative to other states in $^{229}$Pa can be put in
correspondence to similar increments for the states at 831.8 and 938.6 keV relative to the g.s.
in $^{228}$Th. But this state, as well as other low-lying states in $^{229}$Pa, are only weakly
excited in contrast to the strong excitation of the state 831.8 keV in  $^{228}$Th. At the same
time, the $3/2^-$ state at 1523.7 keV is excited strongly, but it cannot be put in
correspondence to the first excited state in $^{228}$Th: there is practically no increment of
its moment of inertia relative to the lowest 3/2$^-$ state. Besides that, its energy is almost
twice larger than the first excited $0^+$ state in $^{228}$Th.

From the insufficient information we can only conclude that the 831.8 keV state in $^{228}$Th
has the largest pairing vibrational component and in $^{229}$Pa the additional proton has the
effect that the largest pairing vibrational component is moved to the 1523.7 keV state. No
theoretical explanation was undertaken since the publication of the experimental results on
$^{229}$Pa \cite{Lev94}. It would be interesting to undertake the theoretical analysis of the
excitations with the $L=0^+$ transfer in odd nuclei and first of all in the $^{229}$Pa nucleus.
Experimental study of excitations in other odd nuclei, as we have seen, may promise unexpected
phenomena.

\begin{figure}
\begin{center}
\epsfig{file=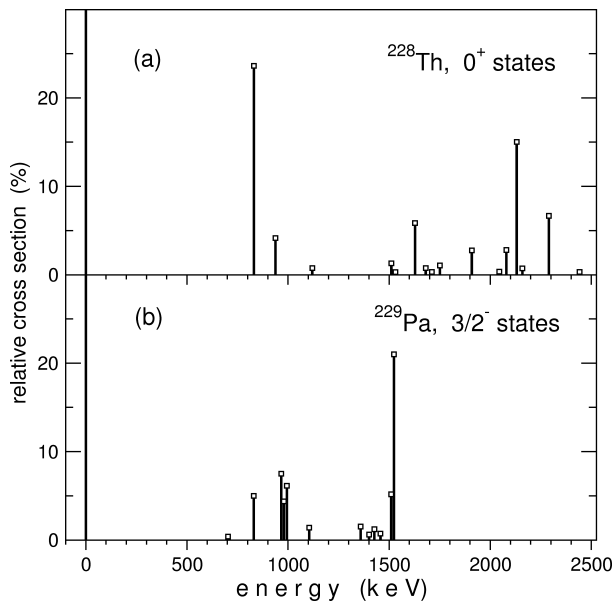, width=8 cm, angle=0} \caption{\label{fig:Th_Pa} Comparison of the (p,t)
cross sections for the $0^+$ states in $^{228}$Th and $3/2^-$ states  in $^{229}$Pa
  relative to  the cross sections for  the ground state in $^{228}$Th and for the lowest
 3/2$^-$ state in $^{229}$Pa (in percent). Energies in $^{229}$Pa are given relative to the
 lowest  3/2$^-$ state.}
\end{center}
\end{figure}

As for the experimental evidence of the nature of other $0^+$ states, we have only the moments
of inertia derived from the sequences of states treated as rotational bands and thus only
tentative conclusions can be drawn about their structure. In contrast to $^{230}$Th
\cite{Lev94},  for which they are distributed almost uniformly over the region from 47 to 98
MeV$^{-1}$, the moments of inertia  in $^{228}$Th have values close to 50 MeV$^{-1}$ only for
the g.s., $\beta$-vibrational $0^+$ states and  for the state at 1531.7 keV, all other $0^+$
states have larger values from 70 to 95 MeV$^{-1}$.
 This fact can  indicate that corresponding states are possibly of two-phonon nature too, or two
quasi-particle states with an admixture of the pairing vibrations.

\section{Conclusion}

Excited states in $^{228}$Th have been studied in (p,t) transfer reactions.  106 levels were
assigned, using a DWBA fit procedure, additionally only the energies are determined for 57
states.  Among them, 17 excited $0^+$ states have been found in this nucleus up to an energy 2.5
MeV, most of them have not been experimentally observed before. Their accumulated strength makes
up for more than 70\,\% of the ground-state strength. Firm assignments have been made for most
of the 2$^+$ and 4$^+$ states and for some of the 6$^+$ states. These assignments allowed to
suggest multiplets of states, which can be treated as one- and two-phonon octupole quadruplets,
and to identify the sequences of states, which have the features of rotational bands with
definite inertial parameters. Moments of inertia are derived from these sequences. Only for the
g.s. and $\beta$-vibrational states and additionally for the state at 1531.7 keV (for which the
shape of the angular distribution  is  different from most other ones), the moments of inertia
are about 50 MeV$^{-1}$. For all other states they are larger than 70 MeV$^{-1}$, i.e. the value
for the first excited $0^+$ state.  This information, together with the spectroscopic
information on some ${\gamma}$-transitions, were used for  conclusions on the nature of the
$0^+$ states. The experimental data have been compared to {\it spdf}-IBM and QPM calculations.
Spectroscopic factors from the (p,t) reaction, and the trend in their change with excitation
energy, are approximately reproduced by both the IBM and QPM for the 0$^+$ states. A remarkable
feature of the IBM and QPM is the prediction of strong first vibrational excitations, close in
magnitude and position to the experimental ones. Giving also an approximately correct number of
$0^+$ states, these models provide different predictions for the structure of these states. The
lack of additional information does not allow for final conclusions on the validity of the
theoretical approaches. Challenging experiments on gamma spectroscopy following (p,t) reactions
would give much needed information.

\section{Acknowledgements}

The work was supported by the
DFG (C4-Gr894/2-3, Gu179/3, Jo391/2-3, Zl510/4-2), MLL, and  US-DOE, contract number
DE-FG02-91ER-40609,  and CZ.1.05/2.1.00/03.0082.



\begin{thebibliography}{99}
\bibitem{Mah72} J.~V.~Maher, J.~R.~Erskine, A.~M.~Friedman,
R.~H.~Siemssen, and J.~P.~Schiffer, Phys. Rev. C {\bf 5}, 1380 (1972).

\bibitem{Dal87} J.~Dalmasso, H.~Maria, and G.~Ardisson, Phys.Rev. C {\bf 36}, 2510 (1987)

\bibitem{Lev94}   A.~I.~Levon,  J.~de Boer, G.~Graw, R.~Hertenberger,
D.~Hofer, J.~Kvasil, A.~L\"osch, E.~M\"uller-Zanotti, M.~W\"urkner, H.~Baltzer,
 V.~Grafen, and C.~G\"unther,  Nucl. Phys.  {\bf A576}, 267 (1994).

\bibitem{Wir04} H.-F.~Wirth, G.~Graw, S.~Christen, D.~Cutoiu, Y.~Eisermann,
C.~G\"unther, R.~Hertenberger, J.~Jolie, A.~I.~Levon, O.~M\"oller, G.~Thiamova, P.~Thirolf,
D.~Tonev, and N.~V.~Zamfir, Phys. Rev. C {\bf 69}, 044310 (2004).

\bibitem{Bal96} H.~Baltzer, J.~de~Boer, A.~Gollwitzer, G.~Graw, C.~Gunther, A.~I.~Levon, M.~Loewe,
               H.~J.~Maier, J.~Manns, U.~M\"uller, B.~D.~Valnion, T.~Weber, M.~W\"urkner,
               Z. Phys. {\bf A356}, 13 (1996).

\bibitem{Lev09} A.~I.~Levon, G.~Graw, Y.~Eisermann, R.~Hertenberger, J.~Jolie,
N.~Yu.~Shirikova, A.~E.~Stuchbery, A.~V.~Sushkov, P.~G.~Thirolf, H.-F.~Wirth, and N.~V.~Zamfir,
Phys. Rev. C {\bf 79}, 014318 (2009).

\bibitem{Web98} T.~Weber, J.~de~Boer, K.~Freitag, J.~Gr\"oger, C.~G\"unther, P.~Herzog,
V.~G.~Soloviev, A.~V.~Sushkov, and N.~Yu.~Shirikova,  Eur. Phys. J.  {\bf A3}, 25 (1998)

\bibitem{Schu86} P.~Schuler, Ch.~Lauterbach, Y.~K.~Agarwal, J.~De~Boer, K.~P.~Blume, P.~A.~Butler, K.~Euler,
Ch.~Fleischmann, C.~G\"unther, E.~Hauber, H.~J.~Maier, M.~Marten-Tolle, Ch.~Schandera,
R.~S.~Simon, R.~Tolle, and P.~Zeyen,  Phys. Lett. {\bf B174}, 241 (1986).

\bibitem{Zan91} E.~Zanotti, M.~Bisenberger, R.~Hertenberger, H.~Kader,
and  G.~Graw, Nucl. Instrum. Methods A {\bf 310}, 706 (1991).

\bibitem{Wir01} H.-F.~Wirth, Ph.~D.~thesis, Techn. Univ. M\"unchen, 2001,
(http://tumb1.biblio.tu-munchen.de/publ/diss/ph/2001/wirth.html).

\bibitem{Rie91} F.~Riess, Beschleunigerlaboratorium M\"unchen, Annual
Report, 1991, p.168.

\bibitem{Bec69} F.~D.~Becchetti and G.~W.~Greenlees, Phys. Rev. {\bf 182},  1190 (1969).

\bibitem{Fly69} E.~R.~Flynn, D.~D.~Amstrong, J.~G.~Beery, and A.~G.~Blair,
                Phys. Rev. {\bf 182}, 1113 (1969).

\bibitem{Bec71} F.~D.~Becchetti and G.~W.~Greenlees, Proc. Third Int. Symp. on
               polarization phenomena in nuclear reactions, Medison, 1970,
               ed. H.~H.~Barshall and W.~Haeberli (University of Wisconsin
               Press, Medison, 1971) p.68

\bibitem{Kun} P.~D.~Kunz, computer code CHUCK3, University of Colorado,
unpublished.

               Acta Phys. Pol. {\bf B29}, 365 (1998).

\bibitem{Bur85} D.~G.~Burke, B.~L.~W.~Maddaock, W.~F.~Davidson, Nucl.Phys.  {\bf A442}, 424 (1985).

\bibitem{Eng87} J.~Engel and F.~Iachello, Nucl. Phys. {\bf A472}, 61 (1987).

\bibitem{Zam01} N.~V.~Zamfir and D.~Kusnezov, Phys. Rev. C {\bf 63}, 054306 (2001).

\bibitem{Zam03} N.~V.~Zamfir and D.~Kusnezov, Phys. Rev. C {\bf 67}, 014305 (2003).

\bibitem{Pas10} S.~Pascu, N.~V.~Zamfir, Gh.~ C\u ata-Danil, and N.~M\u arginean, Phys. Rev. C {\bf 81}, 054321 (2010).

\bibitem{Kuz90} D.~Kusnezov, J. Phys. A {\bf 23}, 5673 (1990).

\bibitem{Kuz89} D.~Kusnezov, J. Phys. A {\bf 22}, 4271 (1989).

\bibitem{Iach87} F.~Iachello and A.~Arima, The Interacting Boson Model (Cambridge University Press, Cambridge, England, 1987).

\bibitem{Pas_unp} S.~Pascu {\it et al.}, unpublished.

\bibitem{Kuz_comp} D.~Kusnezov, computer code OCTUPOLE (unpublished).

\bibitem{LoI04} N.~Lo~Iudice, A.~V.~Sushkov, and N.~Yu.~Shirikova, Phys. Rev. C {\bf 70}, 064316 (2004).

\bibitem{LoI05} N.~Lo~Iudice, A.~V.~Sushkov, and N.~Yu.~Shirikova, Phys. Rev. C {\bf 72}, 034303 (2005).

\bibitem{Sol92} V.~G.~Soloviev,
{\it Theory of Atomic Nuclei: Quasiparticles and Phonons} (Institute of Physics, Bristol, 1992).

\bibitem{Sush} A.~V.~Sushkov, private communication.

\bibitem{Pya67} N.I. Pyatov, Ark. Phys. {\bf 36}, 667 (1967).

\bibitem{Sol76} V.~G.~Soloviev, Theory of Complex Nuclei, Pergamon Press, Oxford (1976).

\bibitem{Gra03}  G.~Graw, Y.~Eisermann, R.~Hertenberger, H.-F.Wirth, S.~Christen, O.~Moeller, D.~Tonev, J.~Jolie,
C.~Guenther, A.~I.~Levon, and N.~V.~Zamfir,  Int. Conf. on Nucl. Reaction Mechanisms, Varenna,
Italy, June 9-13, 2003.

\bibitem{Lev98} A.~I.~Levon, J.~de~Boer, M.~Loewe, M.~W\'rkner, T.~Czosnyka, J.~Iwanicki,
and P.~J.~Napiorkowski, Eur. Phys. J. {\bf A2}, 9 (1998).

\end{thebibliography}
\end{document}